\documentclass[a4paper,11pt]{article}
\usepackage{jcappub}
\usepackage{lineno}
\usepackage{mathrsfs}
\usepackage{booktabs}
\usepackage{amsmath}
\usepackage{xcolor}
\usepackage{float}
 \usepackage{tabularx}
 \usepackage{array}
 \usepackage{soul}
 \usepackage{multirow}
\allowdisplaybreaks
\DeclareMathOperator{\csch}{csch}
\sethlcolor{red!30}

\arxivnumber{2605.20102} 
\title{\boldmath Non-relativistic cosmological collider signals}

\author[a]{Matheus~C.~Ferreira,}
\author[a,b]{Felipe~T.~Falciano}
\affiliation[a]{Brazilian Center for Research in Physics (CBPF), \\ Dr. Xavier Sigaud st. 150, zip 22290-180, Rio de Janeiro, RJ, Brazil}
\affiliation[b]{PPGCosmo, CCE, Federal University of Espírito Santo (UFES), \\
Av. Fernando Ferrari, 540, CEP 29.075-910, Vitória, ES, Brazil}

\emailAdd{matheuscurado@cbpf.br}
\emailAdd{ftovar@cbpf.br}

\abstract{
We investigate a non-relativistic realization of the boostless cosmological collider in a scenario where inflationary interactions are mediated by a massive tilted-ghost spectator field. Unlike standard boostless collider constructions, in which the characteristic non-Gaussian signatures are mainly generated by boost-breaking interaction vertices, the dominant effect in the present framework arises directly from the propagation of the spectator modes. Non-relativistic corrections deform the bulk mode functions, inducing a tilt that modifies the in-in correlators and generates a distinctive collider signal. The resulting squeezed-limit non-Gaussianity reproduces the qualitative structure of boostless cosmological-collider signals while originating from a fundamentally different dynamical mechanism. A central feature of the construction is the emergence of a chemical-potential-like parameter that controls the relative weight of the two late-time oscillatory branches. However, the tilted-ghost mode exhibits distinctive dynamical features and does not correspond to a chemical-potential deformation. 
Depending on the sign of the tilt, the corresponding non-Gaussian signal can be either enhanced or suppressed. We show that the tilted-ghost scenario provides a simple effective framework in which boostless-collider phenomenology and branch asymmetries arise naturally from non-relativistic propagation effects.}

\begin{document}
\maketitle
\flushbottom

\clearpage

\section{Introduction}
\label{sec:intro}

Primordial non-Gaussianity provides a direct window into the microscopic physics operative during inflation. This idea has been developed systematically over the last decade under the banner of ``Cosmological Collider physics''~\cite{Arkani-Hamed:2015bza,
Arkani-Hamed:2018kmz,
Pinol_2023,
Qin_2023,
Qin:2025xct,
Aoki:2023wdc,
Aoki:2024uyi,
Jiang:2025mlm,
Jazayeri:2023xcj,
Jazayeri:2023kji,
Jazayeri:2022kjy,
DuasoPueyo:2023kyh,
Baumann_2012,
Baumann_2018,
Lee_2016,
Wu_2019,
Saito_2018,
Lu_2020,
Lu_2021,
Sou_2021,
Hook_2020,
Hook_20201,
Liu_2020,
Wang_2020,
Fan_2021,
Kumar_2020,
Hubisz:2024xnj,
Cui_2022,
Chua:2018dqh,
Chen_2018,
Chen_2017,
Chen_2012,
Chen:2009zp,
Kehagias_2017,
Gong_2013,
Noumi_2013}.
During the inflationary epoch
\cite{
Guimaraes:2020vfd,
Iacconi:2024hmg,
Falciano:2013uaa,
Mukhanov:1990me,
Maldacena:2002vr,
Martin2014,
Planck:2018jri},
the accelerated expansion stretches the quantum fluctuations of massive fields to superhorizon scales, enabling them to leave non-analytic signatures in primordial correlators before their amplitudes are diluted by the expansion. These signatures are most clearly visible in the squeezed limit of the bispectrum and in the collapsed limit of the trispectrum, where they encode spectroscopic information about the mass and spin of the exchanged state.\footnote{The mass is reflected in the dependence on the ratio between long and short momenta, whereas the spin is encoded in the angular dependence between these modes.} In this sense, inflation acts as a high-energy probe of new degrees of freedom, potentially sensitive to particles with masses at or above the Hubble scale that would otherwise be inaccessible.\footnote{The inflationary Hubble scale remains unknown; a detection of primordial tensor modes would determine it in Planck units. While many inflationary models point to a high scale, phenomenology still allows for much lower values, provided reheating is completed before Big Bang nucleosynthesis.}

Recent developments have extended this program in several complementary directions. On the theory side, these include effective-field-theory bases for heavy sectors coupled to the inflaton, massive spinning fields, compact or light scalar sectors, and new bootstrap or loop representations of cosmological correlators~\cite{Craig:2024wue,Cheung:2025spinning,Chakraborty:2023qgg,Chakraborty:2023zxi,Chakraborty:2025lightcompact,Aoki:2024multiple,Kumar:2025warped,Quintin:2024fingerprints,Jazayeri:2025wrinkle,Graefe:2026split,Colas:2025open}. On the phenomenological side, full-shape computations and direct CMB and large-scale structure searches have started to turn collider templates into observation-oriented observables~\cite{Sohn:2024planck,Suman:2025significance,Suman:2025cmbII,Cabass:2024boss,Goldstein:2024massiveish,Goldstein:2025wiggly,Green:2026extending,Kumar:2026beyond,Kumar:2026scalars}. Higher-point observables provide an additional arena, especially through trispectrum searches and parity-sensitive galaxy-survey observables~\cite{Philcox:2025cmbtrispectrumI,Philcox:2025cmbtrispectrumII,Philcox:2025cmbtrispectrumIII,Bao:2025parity}.

At the same time, in slow-roll inflation and in quasi-single-field scenarios, primordial non-Gaussianities are often small. In addition, the production of a heavy particle\footnote{More precisely, a particle belonging to the principal series of irreducible representations of $SO(1,4)$.} in de Sitter space is subject to the familiar Boltzmann suppression $e^{-\pi\mu}$, with $\mu\sim m/H$. As a result, the contribution of a heavy exchanged state to the squeezed bispectrum is typically exponentially suppressed, even though it retains the characteristic oscillatory dependence on the logarithm of the momentum ratio. This logarithmic behavior encodes the mass of the exchanged particle and gives rise to the interference pattern that underlies the cosmological collider signal.

Several mechanisms have been proposed to soften this suppression, including non--Bunch--Davies initial states~\cite{Yin:2023jlv}, couplings for which the inflaton background acts as a chemical potential~\cite{Bodas_2021,Bodas:2024hih,Bodas:2025chargedloops,Aoki:2026modularCP,Kumar:2026beyond,Chen:2026resonant}, transient tachyonic evolution~\cite{McCulloch:2024tachyon}, and additional interaction structures designed to enhance the non-Gaussian signal~\cite{Pimentel:2022fsc,Qin:2023ejc,Wang:2025qww,deRham:2025bispectrumIslands}. In previous work, it was shown that a ghost spectator can already amplify the cosmological collider signal relative to the standard de Sitter case~\cite{Ferreira:2026ghostspectators}. More recently, the broader viewpoint of the ``boostless collider'' has further emphasized that inflation itself breaks Lorentz boosts, so collider-like signatures need not be tied exclusively to the relativistic de Sitter template.

Motivated by these developments, in this work we consider spectator modes inspired by tilted ghost inflation \cite{Senatore:2004rj} and investigate their role in cosmological collider observables. Our goal is not to embed the full collider setup in tilted ghost inflation as a background theory, but rather to use a massive tilted-ghost-inspired spectator as the exchanged mediator of inflationary interactions. The main idea is to understand how non-relativistic corrections modify the spectrum of the collider signal. In our setup, these corrections deform the spectator mode functions, induce a tilt in the corresponding modes, and thereby reshape the in-in correlators. This provides a non-relativistic realization of boostless-collider phenomenology with a distinct dynamical origin: instead of relying primarily on new boost-breaking interaction vertices, the signal is modified already at the level of propagation of the exchanged spectator.

It is useful to contrast the present setup with the boostless collider in the presence of a sound-speed hierarchy \cite{Pimentel:2022fsc,Jazayeri:2023xcj,Jazayeri:2022kjy,Green:2026extending}. In that case, the new effect appears primarily in the argument of the oscillatory signal. Schematically, one has \footnote{Here $S$ denotes the dimensionless squeezed-limit shape function, evaluated in the regime $k_L\equiv k_3\ll k_1\simeq k_2\equiv k_S$, and the expressions below display only the leading non-analytic clock contribution.}
\begin{equation}
\boxed{\begin{aligned}
\text{Boostless Speed of Sound}
&\;\Longrightarrow\;
S_{\text{\rm non\mbox{-}an}}
\;\sim\;
\left(\frac{k_3}{k_1}\right)^{1/2}
\cos\!\left[
\mu \log\!\left(c_{b}^2\frac{k_3}{k_1}\right)
+\delta(\mu)
\right]
\end{aligned}}
\end{equation}
so that the sound-speed ratio shifts the logarithmic clock argument. By contrast, in our non-relativistic boostless setup, the relevant parameter is the signed tilt $\lambda = s_{\rm tilt}\beta/(4\gamma)$, where $s_{\rm tilt}=\pm$ corresponds to the negative and positive tilt, generated directly by the tilted spectator mode functions. In this case, the collider signal is modified both in amplitude and in phase,
\begin{equation}
\boxed{\begin{aligned}
\text{Non-Relativistic Boostless}
\;\Longrightarrow\;
S_{\text{\rm non\mbox{-}an}}
\sim
\left(\frac{k_3}{k_1}\right)^{1/2} 
\cos\!\left[
\mu \log\!\left(\gamma \frac{k_3}{k_1}\right)
+\delta_{\rm TG}(\mu,\lambda,\gamma)
\right]
\end{aligned}}
\label{eq:intro_TG_template}
\end{equation}
Thus, whereas the boostless sound-speed effect mainly reorganizes the argument of the oscillation, the non-relativistic boostless effect associated with tilted modes simultaneously dresses the size and phase of the signal.

A useful way to summarize the picture is to contrast three situations: the usual relativistic de Sitter collider, the boostless collider, and the present non-relativistic spectator realization. In the de Sitter case, the signal is controlled by the standard horizon crossing condition and suffers the usual Boltzmann suppression. In boostless scenarios, the signal can be modified by the breaking of relativistic boost symmetry, often through sound-speed effects or new interaction structures. In the present setup, by contrast, the modification arises because the exchanged massive spectator has non-relativistically corrected mode functions, whose tilt acts as an effective branch-asymmetry in the late-time connection coefficients. This makes the massive tilted-ghost spectator a natural mediator of collider-like signals beyond the standard relativistic paradigm.

The relevant spectator sector is inspired by the ghost condensate, where Lorentz invariance is spontaneously broken~\cite{Hamed_2004, Nima_Arkani_Hamed_2004}. In the ghost limit, the corresponding scalar fluctuation obeys a modified dispersion relation of the schematic form $\omega^2\propto k^4$, characteristic of higher-derivative operators. In the tilted case, the mode functions are normalized so as to recover the ghost-condensate mode in the early-time regime. The non-relativistic corrections then deform this structure further and induce a tilt in the massive spectator modes. This tilt has an important physical consequence for collider phenomenology: it acts as a branch-asymmetry parameter in the mode functions and, through the Schwinger--Keldysh computation, biases the relative weight of the two late-time oscillatory branches in close analogy with chemical-potential mechanisms previously discussed in the cosmological collider literature. Depending on the sign of the tilt, the signal can be enhanced or suppressed relative to the pure-ghost reference branch.

\begin{table}[ht]
\centering
\setlength{\tabcolsep}{4pt}
\resizebox{\textwidth}{!}{%
\begin{tabular}{lcccc}
\toprule
 & \textbf{dS} & \textbf{Boostless} & \textbf{Ghost} & \textbf{Tilted ghost} \\
\midrule
UV limit
& relativistic BD
& $c_b k$ BD-like
& $k^2$ WKB
& tilted $k^2$ WKB \\[0.6em]
Clock phase
& $\mu\log(k_L/k_S)+\delta$
& $\mu\log(c_b^2 k_L/k_S)+\delta$
& $\mu\log(\gamma k_L/k_S)+\delta$
& $\mu\log(\gamma k_L/k_S)+\delta(\lambda)$ \\[0.6em]
\multirow{2}{*}{Amplitude effect}
& Boltzmann & sound-speed & pure-ghost & $\lambda$-dependent \\
& suppressed & dependent & normalization &enhancement \\[0.6em]
\multirow{2}{*}{Physical origin}
& massive dS & boost-breaking & quartic & tilted \\
& exchange & vertices &dispersion & propagation \\
\bottomrule
\end{tabular}%
}
\caption{Qualitative comparison between the standard de Sitter, boostless, ghost, and tilted-ghost spectator scenarios. The entries identify the physical origin of each effect rather than universal numerical freeze-out formulae. In the tilted-ghost case the quadratic correction changes the mode evolution through a crossover and biases the late-time Whittaker connection coefficients. The sign of the signed tilt parameter determines whether the corresponding clock envelope is enhanced or suppressed.}
\label{tab:ds_boostless_ghost}
\end{table}

To make these statements precise, we construct the relevant bulk-to-bulk and bulk-to-boundary propagators within the Schwinger--Keldysh formalism \cite{Chen:2017ryl, Weinberg_2005, calzeta,Colas:2025open,Graefe:2026split}. We then study the exchange bispectrum generated by inflaton fluctuations interacting through a massive tilted-ghost spectator. The non-relativistic dispersion relation modifies the structure of the mode functions and hence the corresponding in-in integrals.~As a consequence, the leading non-analytic contributions persist but with modified scaling behavior, a tilted phase structure, and a sign-dependent envelope. In the squeezed limit, we show explicitly that the characteristic oscillatory features remain present while the amplitude can be enhanced or suppressed by the tilted-ghost connection coefficients.

\vskip 4 pt
\noindent {\bf Outline}: This paper is organized as follows. In Section~\ref{sec:model}, we introduce the tilted-ghost spectator and review the relevant mode functions, including the crossover structure and the late-time Whittaker branches. In Section~\ref{sec:collider}, we formulate the exchange bispectrum in the Schwinger--Keldysh formalism and isolate the non-analytic soft contribution. In Section~\ref{sec:phenomenology}, we analyze the resulting squeezed-limit signal and its dependence on the effective chemical-potential parameter. The long Schwinger-Keldysh bookkeeping, the monomial time integrals, the hard-channel expansion, and the Whittaker late-time propagator calculation are collected in the appendices.

\noindent {\bf Notation and Conventions}: We work with the metric signature \( (-,+,+,+) \). Throughout the paper, we use natural units,
\( \hbar = c = 1 \). Scalar fields associated with the inflaton sector are denoted by
\( \varphi \) and \( \phi \), while ghost fields are denoted by \( \sigma \).
Spacetime indices are labeled by \( \alpha = (\eta, 1, 2, 3) \), where \( \eta \) denotes conformal
time, which is used throughout this work. Latin indices \( i = 1,2,3 \) refer to spatial
components. Three dimensional vectors are written in boldface, \( \mathbf{k} \). The magnitude of a vector
is defined as \( k = |\mathbf{k}| \), and the corresponding unit vector is
\( \hat{\mathbf{k}} = \mathbf{k}/k \). The momentum of the \( n \)-th external leg of a correlation
function is denoted by \( \mathbf{k}_{n} \), with magnitude \( k_{n} \equiv |\mathbf{k}_{n}| \). Primed correlators denote stripped correlators, with the overall momentum-conserving delta function removed, e.g. \(\langle \mathcal O_{\mathbf k_1}\cdots \mathcal O_{\mathbf k_n}\rangle=(2\pi)^3\delta^{(3)}(\sum_i\mathbf k_i)\langle \mathcal O_{\mathbf k_1}\cdots \mathcal O_{\mathbf k_n}\rangle'\).

\section{The Model}
\label{sec:model}

In our setup, gravitational dynamics is described by the Einstein--Hilbert action. The inflationary background is driven by a standard slow-roll inflaton field, while a non-relativistic tilted-ghost scalar is introduced as a spectator degree of freedom coupled to the inflaton sector. Since the spectator field contributes negligibly to the total energy density, it does not affect the background evolution, which remains well approximated by a quasi-de Sitter spacetime. Nevertheless, the modified propagation of the tilted-ghost field and its interactions with the inflaton leave distinctive imprints on the primordial correlation functions generated during inflation.

The framework considered here should be viewed as an effective description inspired by tilted ghost inflation. Rather than embedding the cosmological-collider setup into the full tilted-ghost background, we retain the standard inflationary dynamics and incorporate the tilted-ghost dispersion relation as a modification of the spectator sector. This description is valid provided the spectator energy density remains subdominant, allowing backreaction on the inflationary background to be neglected and the non-relativistic dispersion relation to be treated within a controlled EFT expansion. In particular, the parameter $\gamma=H/M$ characterizes the hierarchy between the Hubble scale and the ghost scale, while the signed tilt parameter $\lambda=s_{\rm tilt}\beta/(4\gamma)$ quantifies deviations from the pure ghost-condensate limit. A complete UV completion of the spectator sector and its coupling to the inflationary background lies beyond the scope of the present work and will be investigated elsewhere. 

We therefore consider a minimal setup consisting of the inflaton field $\phi$ and a massive spectator scalar $\sigma$. The inflaton fluctuation is described by the quantum field $\varphi(\mathbf{x},\eta)$, whose Fourier expansion is given by

\begin{equation}
\varphi_{\mathbf{k}}(\eta)
=
u(k,\eta)\, b_{\mathbf{k}}
+
u^{*}(k,\eta)\, b^{\dagger}_{-\mathbf{k}},
\label{phimodeexp}
\end{equation}
where $b_{\mathbf{k}}|0\rangle=0$, and the initial condition for the mode function $u(k,\eta)$ is chosen according to the Bunch-Davies prescription.
At zeroth order in slow roll, the inflationary background is well approximated by de Sitter spacetime
\cite{Baumann:2022mni,Creminelli:2012qr,Mukhanov:1990me,Linde:1981mu,Maldacena:2002vr}.
For a scalar field of mass $m$ on de Sitter, the corresponding mode equation is
\begin{equation}
\left[
\partial_{\eta}^{2}
-
\frac{2}{\eta}\partial_{\eta}
+
k^{2}
+
\frac{m^{2}}{(H\eta)^{2}}
\right]
u(k,\eta)
=
0.
\label{inflatonmodeeq}
\end{equation}
It is convenient to introduce the parameter
\begin{equation}\label{eq:nu_mu_def}
\nu \equiv \sqrt{\frac{9}{4}-\frac{m^{2}}{H^{2}}} 
\quad ,
\end{equation}
and the Bunch--Davies solution can be written as
\begin{equation}
u(k,\eta)
=
\frac{H\sqrt{\pi}}{2\,k^{3/2}}\,
e^{i\pi(\nu+\frac12)/2}\,
(-k\eta)^{3/2}
\mathcal{H}^{(1)}_{\nu}(-k\eta).
\label{modeinfm}
\end{equation}
Two special cases can play a particularly important role in this type of scenario. The first corresponds to the conformally coupled scalar field, $m^{2}=2H^{2}$ ($\nu = 1/2$), while the second corresponds to the massless inflaton fluctuation, $m=0$ ($\nu = 3/2$). In both cases, the mode functions simplify to
\begin{equation}
u(k,\eta)=
\begin{cases}
-\dfrac{H\eta}{\sqrt{2k}}\,e^{-ik\eta}
&  ,\quad \text{conformally coupled $(m^{2}=2H^{2})$} \\ 
&\\
\dfrac{H}{\sqrt{2k^{3}}}\,(1+ik\eta)\,e^{-ik\eta}

& , \quad \text{massless inflaton $(m=0)$}
\end{cases}
\label{modeinfm0}
\end{equation}
The tilted-ghost mode equation was originally investigated in Ref.~\cite{Senatore:2004rj} within a WKB approximation scheme. As we will demonstrate below, this equation also admits an exact analytic solution, which proves to be more suitable for the purposes of the present analysis.

\subsection{Non-Relativistic Corrections}
\label{sec:TG_modes}

The main effect of the non-relativistic correction is to deform the spectator-field mode functions already at the level of the bulk propagation, thereby modifying both the amplitude and the phase of the late-time signal. To investigate this effect, we consider a massive scalar spectator field propagating on a de Sitter background,
\begin{equation}
a(\eta)=-\frac{1}{H\eta},
\qquad \eta<0,
\label{eq:scale_factor}
\end{equation}
and work with the canonically normalized variable $f_k(\eta)\equiv a(\eta)\,\sigma_k(\eta)$. For an ordinary massive spectator in de Sitter, $f_k$ satisfies
\begin{equation}
f_k''(\eta)
+
\left[
k^2-\frac{\nu^2-\tfrac14}{\eta^2}
\right]f_k(\eta)=0,
\label{eq:standard_massive_f}
\end{equation}
with $\nu$ defined in \eqref{eq:nu_mu_def}. For fields in the principal series, we parameterize $\nu=i\mu$, with $\mu>0$. The tilted-ghost spectator is then obtained by replacing the relativistic dispersion relation according to
\begin{equation}
k^2
\;\longrightarrow\;
\pm\,\beta k^2+\gamma^2 k^4\eta^2,
\qquad
\gamma\equiv \frac{H}{M},
\qquad \beta>0,
\label{eq:TG_dispersion}
\end{equation}
where the $\pm$ sign corresponds to the two possible tilt choices introduced by Senatore in Ref.~\cite{Senatore:2004rj}. Here $M$ denotes the ghost scale suppressing the higher-derivative operator responsible for the quartic dispersion, so that $\gamma=H/M$ measures the size of the ghost correction in Hubble units. With this convention, the equation of motion takes the form
\begin{equation}
f_k''(\eta)
+
\left[
\pm\,\beta k^2
+\gamma^2 k^4\eta^2
-\frac{\nu^2-\tfrac14}{\eta^2}
\right]
f_k(\eta)=0.
\label{eq:TG_mode_equation}
\end{equation}
The parameter $\gamma$ controls the strength of the ghost contribution, while $\beta$ characterizes the magnitude of the relativistic correction, with its sign determining the choice of tilt. Before solving Eq.~\eqref{eq:TG_mode_equation}, it is useful to identify the asymptotic far-past regime. In the limit $|\eta|\to\infty$, the ghost contribution becomes dominant, and the mode equation simplifies to
\begin{equation}
f_k''(\eta)+\gamma^2k^4\eta^2\,f_k(\eta)\simeq 0.
\end{equation}
The corresponding WKB frequency is $\omega_k(\eta)\simeq \gamma k^2 \eta$ and the associated positive-frequency solution therefore takes the form,
\begin{equation}
f_k(\eta\to-\infty)=
\frac{1}{\sqrt{2\omega_k(\eta)}}
\exp\!\left(-i\int^\eta \omega_k(\tilde\eta)\,d\tilde\eta\right)
=
\frac{e^{-\tfrac{i}{2}\gamma k^2\eta^2}}
{k\,\sqrt{-2\gamma \eta}}.
\label{eq:early_time_normalization_clean}
\end{equation}
We normalize the exact solution by requiring that it reproduces \eqref{eq:early_time_normalization_clean} in the asymptotic past. An exact solution to Eq.~\eqref{eq:TG_mode_equation} can be obtained by introducing the variable $z=i\gamma k^2\eta^2$ and expressing the mode function as $f_k(\eta)=g(z)$. With this change of variables, Eq.~\eqref{eq:TG_mode_equation} takes the form
\begin{equation}
z\,g''(z)+\frac12 g'(z)
+
\left[
-\frac{z}{4}
\mp\,\frac{i\beta}{4\gamma}
-\frac{\nu^2-\tfrac14}{4z}
\right]g(z)=0.
\label{eq:g_equation}
\end{equation}
Performing one final change of variables, namely defining $g(z)=z^{-1/4}h(z)$, we arrive at
\begin{equation}\label{eq:whittaker_form}
h''(z)
+
\left[
-\frac14
\mp\,\frac{i\beta}{4\gamma}\,\frac{1}{z}
+\frac{1-\nu^2}{4z^2}
\right]h(z)=0 
\quad ,
\end{equation}
which corresponds to the Whittaker differential equation. To avoid sign ambiguities, we distinguish the positive magnitude of the tilt parameter from the signed quantity entering the Whittaker index. We therefore set
\begin{equation}
\lambda_0 \equiv \frac{\beta}{4\gamma}>0,
\qquad
\lambda\equiv s_{\rm tilt}\lambda_0,
\label{eq:lambda_eff_def}
\end{equation}
with
\begin{equation}
\begin{array}{c|c|c|c}
\text{branch} & s_{\rm tilt} & \text{quadratic term in }Q(\eta) & \lambda \\
\hline
\text{negative tilt} & +1 & +\beta k^2 & +\beta/(4\gamma) \\
\text{positive tilt} & -1 & -\beta k^2 & -\beta/(4\gamma)
\end{array}
\label{eq:tilt_sign_table}
\end{equation}\\
Accordingly, the parameter $\lambda$ will always denote the signed effective tilt parameter. The general solution is given by a linear combination of the Whittaker-$W$ and Whittaker-$M$ branches. Imposing the early-time condition \eqref{eq:early_time_normalization_clean} uniquely selects the Whittaker-$W$ branch and fixes the overall normalization. With the signed convention introduced above, the properly normalized canonical mode function is given by
\begin{equation}
f_k(\eta)
=
-\frac{e^{-\pi\lambda/2}}
{\sqrt{2}\sqrt{\gamma}\,k\,(-\eta)^{1/2}}
\,
\mathrm{W}_{- i \lambda,\,\nu/2}
\!\left(i\gamma k^2\eta^2\right).
\label{eq:TG_mode_normalized_compact}
\end{equation}
The positive-tilt solution is therefore obtained by taking $\lambda<0$, whereas the negative-tilt solution corresponds to $\lambda>0$. The sign of the Whittaker argument determines the Stokes wedge selected by the early-time $i\epsilon$ prescription. Using the large-argument asymptotic behavior,
\begin{equation}
W_{\kappa,\mu}(z)\sim e^{-z/2}z^\kappa,\qquad |z|\to\infty,
\end{equation}
one can verify that the branch selected in Eq.~\eqref{eq:TG_mode_normalized_compact} reproduces the WKB phase appearing in Eq.~\eqref{eq:early_time_normalization_clean}, up to an overall physically irrelevant constant phase. Choosing the opposite Stokes wedge would instead complex-conjugate the mode function, corresponding to the opposite frequency prescription. We now extract the late-time behavior in the principal series, corresponding to $\nu=i\mu$. Expanding Eq.~\eqref{eq:TG_mode_normalized_compact} in the limit $\eta\to0^-$, the canonical mode function decomposes into two independent branches,
\begin{equation}
\lim_{\eta\to0^-}f_k(\eta)\sim 
\widetilde{\mathcal A}_+(k,\mu,\beta,\gamma)\,(-\eta)^{\frac12+i\mu}
+
\widetilde{\mathcal A}_-(k,\mu,\beta,\gamma)\,(-\eta)^{\frac12-i\mu},
\label{eq:f_late_time_branches}
\end{equation}
with coefficients fixed by the Whittaker connection formulas. The signed parameter $\lambda=s_{\rm tilt}\beta/(4\gamma)$ enters these coefficients directly, thereby shaping both the amplitude and the phase of the late-time mode. It is straightforward to obtain the physical spectator field $\sigma_k(\eta)$, recalling that 
\begin{equation}
\sigma_k(\eta)=\frac{f_k(\eta)}{a(\eta)}=-H\eta\,f_k(\eta),
\label{eq:sigma_physical}
\end{equation}
one finds the late-time two-point function at unequal times $\eta_{1,2}$,
\begin{align}
\lim_{k\to0}
\langle \sigma_{\mathbf k}(\eta_1)\sigma_{-\mathbf k}(\eta_2)\rangle
&=
\frac{H^2}{2e^{\pi\lambda}}
(k^2\gamma)^{-i\mu}
\frac{(-\eta_1)^{\Delta_-}(-\eta_2)^{\Delta_-}\Gamma(i\mu)^2}
{
\Gamma\!\left(\frac12-i\lambda+\frac{i\mu}{2}\right)
\Gamma\!\left(\frac12+i\lambda+\frac{i\mu}{2}\right)
}
+{\rm c.c.}
\nonumber\\
&\quad+
\frac{H^2}{2\mu\,e^{\pi(\lambda+\mu/2)}}
(-\eta_1)^{\Delta_-}(-\eta_2)^{\Delta_+}
\csch(\pi\mu)
\cosh\!\left(\pi\lambda-\frac{\pi\mu}{2}\right)
\nonumber\\
&\quad+
\frac{H^2}{2\mu\,e^{\pi(\lambda-\mu/2)}}
(-\eta_1)^{\Delta_+}(-\eta_2)^{\Delta_-}
\csch(\pi\mu)
\cosh\!\left(\pi\lambda+\frac{\pi\mu}{2}\right).
\label{eq:sigma_late_time_branches}
\end{align}
The first line exhibits the pair of complex-conjugate late-time powers associated with the oscillatory Whittaker branches. The mixed-power terms correspond to complementary soft-channel contributions. Although these terms originate from local or contact-like structures, the subsequent Schwinger-Keldysh time integrals can still contribute to the coefficients multiplying the same squeezed-limit clock signals. This behavior should be carefully distinguished from the genuinely analytic hard-channel background discussed below.
 
The initial condition for these modes is determined by the pure ghost solution, since in the asymptotic past the term $\gamma^{2} k^{4} \eta^{2}$ dominates the dispersion relation over both the tilt contribution and the mass term. Consequently, the early-time vacuum is selected by the pure ghost regime. This highlights the fact that the spectator field does not admit an exact de Sitter limit in the conventional sense: its ultraviolet behavior is governed by the non-relativistic ghost contribution, and the tilted mode should therefore be interpreted as a deformation of the pure ghost solution rather than of the standard massive de Sitter mode. The tilt modifies the mode function throughout the bulk evolution, and this deformation is subsequently encoded in the late-time coefficients entering the collider observables. In this sense, the non-relativistic correction does not merely introduce an additional interaction vertex, but instead alters the exchanged spectator field already at the level of its propagation. Several limiting regimes of this solution will be investigated in the following sections.

Following Senatore~\cite{Senatore:2004rj}, the tilted-ghost dispersion relation admits two nonequivalent sign choices for the tilt deformation. These correspond to the positive-tilt and negative-tilt cases, distinguished by the sign of the coefficient multiplying the relativistic correction term $\beta k^{2}$. In the asymptotic past, both cases reduce to the same behavior, since the pure ghost contribution $\gamma^{2} k^{4} \eta^{2}$ dominates the dispersion relation and uniquely determines the initial condition. Away from this regime, however, the effect of the tilt becomes considerably more subtle, and its physical implications are not fully captured by the exact solution alone. A proper understanding requires analyzing the turning-point structure of the mode equation, as this structure governs the transition between the different dynamical regimes. We shall now investigate this issue.

\subsection{Dynamics and Crossover}
\label{sec:turning_points}

We organize the dynamics of the tilted-ghost mode equation according to its semiclassical (WKB) regimes. Recall that $\beta>0$, while the tilt is distinguished by the sign choice associated with the relativistic term, namely,
\begin{equation}
f_k''(\eta)+Q_{\pm}(\eta)\,f_k(\eta)=0,
\qquad
Q_{\pm}(\eta)=
\pm\beta k^2+\gamma^2k^4\eta^2-\frac{\nu^2-1/4}{\eta^2},    
\end{equation}
The positive-tilt case is obtained from the negative-tilt case through the replacement $\beta \to -\beta$. In both cases, $Q(\eta)$ acts as an effective squared frequency: the region $Q(\eta)>0$ corresponds to oscillatory evolution, whereas $Q(\eta)<0$ gives rise to evanescent (tachyonic) behavior. Since the inflationary contour is restricted to $\eta\in(-\infty,0)$, the physically relevant turning times correspond to negative real roots. A first general observation is that, at sufficiently early times, the dynamics is universally dominated by the dispersive term.
\begin{equation}\label{eq:early_time}
\eta\to-\infty:
\quad 
Q_\pm(\eta)\simeq \gamma^2k^4\eta^2>0
\quad ,
\quad \omega_k(\eta)\equiv\sqrt{Q_\pm(\eta)}\simeq \gamma k^2|\eta|
\quad .
\end{equation}
Deep in the ultraviolet regime, the evolution is adiabatic and oscillatory, as the effective frequency is dominated by the $\gamma^{2}k^{4}\eta^{2}$ term, thereby fixing a positive-frequency initial condition for either sign of the tilt. The subsequent dynamics are controlled by the competition among the different contributions in $Q_\pm(\eta)$. Before the turning-point region, a crossover occurs between the $k^{4}$ and $k^{2}$ terms, marking the onset of the regime in which the relativistic contribution becomes dynamically relevant. Since the mode function is primarily determined near freeze-out, the tilt leaves an observable imprint only if the $k^{2}$ term dominates prior to freeze-out. Taking our solution in Eq.~(\ref{eq:sigma_physical}) for a massless field $(m=0)$ and evaluating it at late times, we obtain the power spectrum
\begin{equation}
\frac{k^3}{2\pi^2}
\lim_{\eta\to0}
\langle\sigma_k(\eta)\sigma_{-k}^{*}(\eta)\rangle
 = \frac{H^2}{4\pi^2\beta^{3/2}}\quad ,
 \qquad \nu=\tfrac{3}{2}\quad .
\end{equation}
To connect with previous results in the literature, note that the parameter redefinition $\beta \to \beta \delta^{2}$ exactly reproduces the negative-tilt ghost power spectrum obtained in \cite{Senatore:2004rj}. This agreement reflects the hierarchy of scales encoded in the exact solution: the mode is normalized to the adiabatic positive-frequency branch in the ultraviolet, while its late-time amplitude is determined in the regime where the $k^{2}$ term dominates prior to freeze-out. The exact normalization, therefore, does not independently select the freeze-out regime but rather tracks the branch interpolating between the UV $k^{4}$-dominated phase and the late-time $k^{2}$-dominated regime.

For the positive tilt, the dispersion relation acquires the opposite sign, so the $\beta k^2$ term ceases to be stabilizing and instead drives the mode toward a tachyonic phase. This does not obstruct the definition of the vacuum, however, since deep in the ultraviolet regime the $\gamma^{2}k^{4}\eta^{2}$ term dominates the effective frequency, and the evolution remains adiabatic and oscillatory. As the mode redshifts, a crossover occurs at $\eta_{\times}$, where the $k^{4}$ and $k^{2}$ contributions become comparable, namely $|\eta_{\times}|\sim {\sqrt{\beta}}/{\gamma k}$. This marks the onset of the regime in which the relativistic term becomes dynamically relevant, although the crossover should not be identified with the WKB turning point itself. For the light positive-tilt branch, the true turning point is determined by the condition $Q_+(\eta_t)=0$, yielding 
\begin{equation}
\eta_t^2=
\frac{\beta+\sqrt{\beta^2+\gamma^2(4\nu^2-1)}}{2\gamma^2 k^2}.
\label{eq:light_turning_point}
\end{equation}
Around and after the crossover, the $-\beta k^{2}$ term becomes dynamically important and triggers a transient instability. The subsequent de Sitter expansion eventually freezes the mode, interrupting the growth. Hence, the effect of the positive tilt is not to alter the ultraviolet vacuum but rather to introduce an intermediate unstable phase between the UV $k^{4}$-dominated regime and the final frozen configuration. Evaluating the exact solution at late times, we obtain
\begin{equation}
    \frac{k^3}{2\pi^2}\lim_{\eta\to0}\sigma_k(\eta)\sigma_{-k}^{*}(\eta)\bigg|_{-\beta}
    = \frac{H^2\,e^{\pi\beta/2\gamma}}{4\pi^2\beta^{3/2}},
    \qquad \nu=\tfrac{3}{2}.
\end{equation}
After the appropriate rescaling of $\beta$, the positive-tilt regime has a clear physical interpretation. Any convention-dependent overall phase of the mode drops out of the power spectrum. What survives in $\sigma_k\sigma_{-k}^{*}$ is the real exponential enhancement, which carries the physical memory of the transient unstable $-k^2$ interval. In this sense, the crossover region is crucial: it determines whether the mode actually experiences the tachyonic phase before freeze-out. If it does, the late-time power spectrum retains the imprint of that intermediate instability through the normalization of the frozen mode. Thus, the observable effect of the positive tilt is not a residual phase, but the remnant of an unstable epoch occurring after the UV adiabatic regime and before freeze-out. Equivalently, the positive tilt leaves an observable imprint through the exponential enhancement accumulated while the mode crosses the temporary $-k^2$ instability before freeze-out. The phase information is convention-dependent and unobservable in the power spectrum, whereas the enhancement of the late-time amplitude is a genuine physical effect.

\begin{figure}[t]
    \centering
\includegraphics[width=0.75\linewidth,height=6.5cm]{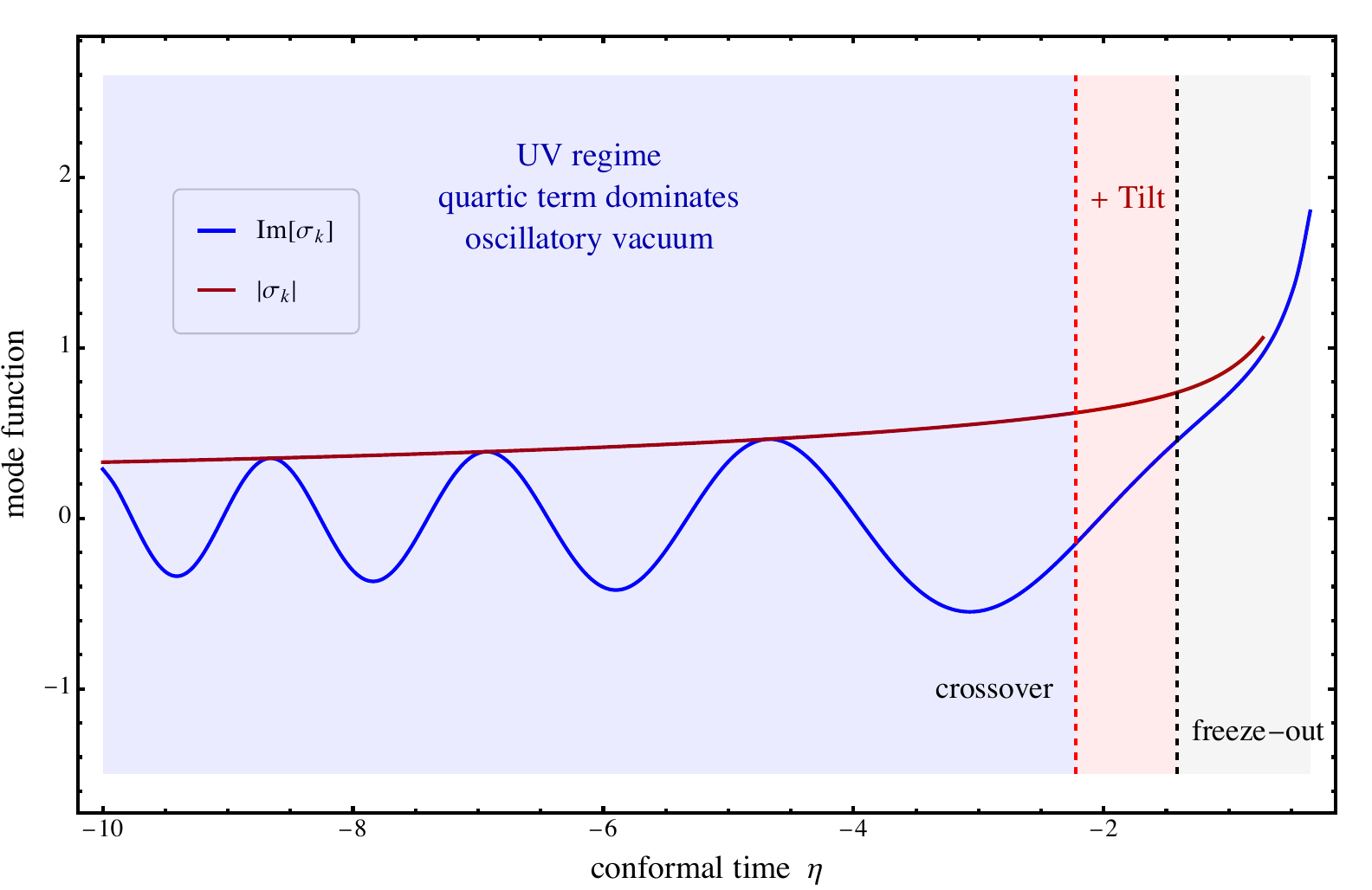}
\caption{Exact positive-tilt mode evolution in conformal time for $k=1$, $\beta=1$, $\gamma=0.45$, and $\nu=3/2$. The blue curve is $\mathrm{Im}\,\sigma_k(\eta)$, while the red curve is the envelope $|\sigma_k(\eta)|$. At early times the quartic term dominates, so the mode is adiabatic and oscillatory. The first dashed line marks the crossover scale $\eta_{\times}\simeq-2.22$, where the quartic and quadratic contributions become comparable. The true WKB turning point lies slightly to the left of the plotted crossover scale; after this point the wrong-sign quadratic term associated with the positive tilt participates in a transient unstable interval, visible as the growth of the envelope. The second dashed line marks the quadratic--mass balance scale $\eta_{\rm fr}\simeq-1.41$, after which the mode approaches its late-time frozen behavior. The shaded regions indicate the corresponding
dynamical regimes.}
\label{positivetiltfig}
\end{figure}

To make this hierarchy of scales explicit, Fig.~\ref{positivetiltfig} displays the exact positive-tilt mode function. The actual WKB turning point, obtained from Eq.~\eqref{eq:light_turning_point}, occurs slightly earlier at $\eta_t\simeq -2.54$. The figure emphasizes the crossover and freeze-out scales because they clearly exhibit the hierarchy governing the mode evolution: an initial oscillatory UV-regime, an intermediate interval in which the negative quadratic contribution becomes dynamically significant, and finally the approach to the frozen late-time configuration. The blue curve represents $\mathrm{Im}(\sigma_k)$, while the red curve corresponds to the amplitude $|\sigma_k|$. The oscillatory behavior of $\mathrm{Im}(\sigma_k)$ reflects the adiabatic ultraviolet vacuum, whereas the mild enhancement of the amplitude accumulates in the interval between $\eta_{\times}$ and $\eta_{\rm fr}$. The observable imprint of the positive tilt therefore does not originate from an overall phase shift of the mode function but rather from the enhancement of its late-time normalization generated during the transient unstable phase preceding freeze-out.

The discussion above was restricted to the light case, $\nu=3/2$, for which the positive-tilt branch may develop a transient unstable interval between the crossover and freeze-out scales. It is instructive to contrast this behavior with the heavy regime, in which the same notion of crossover persists, although its dynamical interpretation changes qualitatively. Indeed, for the heavy regime, we may write $\nu=i\mu$ with $\mu>0$, and the effective frequency reads
\begin{equation}\label{eq:Qplus_heavy}
Q_{\pm}(\eta)=
\pm \beta k^2+\gamma^2k^4\eta^2+\frac{\mu^2+1/4}{\eta^2}
\quad .
\end{equation}
Thus, unlike the light case, the $1/\eta^2$ contribution is now positive. Consequently, the late-time dynamics are no longer driven toward a tachyonic freeze-out. Instead, the heavy mode remains oscillatory in the limit $\eta\to0^{-}$. The crossover scale is still defined by the balance between the quartic and quadratic contributions, which yields $|\eta_{\times}|\sim {\sqrt{\beta}}/{\gamma k}$. The existence of this crossover, however, does not by itself imply the presence of a real turning point. For the heavy positive-tilt branch, a genuine evanescent interval will not form unless the negative quadratic contribution becomes sufficiently large to overcome the positive minimum generated by the remaining terms in Eq.~\eqref{eq:Qplus_heavy}. Therefore, we have the following condition
\begin{equation}
\beta < 2\gamma\sqrt{\mu^2+\frac14},
\qquad \text{or equivalently} \qquad
2|\lambda|<\sqrt{\mu^2+\frac14}.
\label{eq:heavy_no_turning_condition}
\end{equation}
In the controlled small-tilt regime, where the spectator sector is consistently interpreted as a perturbative deformation of the pure ghost mode, this condition is naturally satisfied. The positive contribution $(\mu^{2}+1/4)/\eta^{2}$ then maintains the positivity of the effective frequency along the physical contour, preventing the emergence of a real evanescent interval. Outside this perturbative regime, however, a sufficiently large positive-tilt deformation can drive the minus-sign quadratic term to dominate over the positive barrier, thereby generating a genuine turning region even for principal-series modes. 

This structure is also encoded in the exact Whittaker solution, Eq.~\eqref{eq:TG_mode_normalized_compact}. Its late-time expansion, Eq.~\eqref{eq:sigma_late_time_branches}, shows that in the heavy regime the mode remains oscillatory near the boundary, with the two independent branches weighted by the connection coefficients $A_\pm$. A useful diagnostic of the resulting branch asymmetry is therefore the ratio
\begin{equation}
\left|\frac{A_+}{A_-}\right|^2
=
\frac{
\cosh\!\left[\pi\left(\lambda-\frac{\mu}{2}\right)\right]
}{
\cosh\!\left[\pi\left(\lambda+\frac{\mu}{2}\right)\right]
}\quad .
\label{eq:chemical_potential_ratio}
\end{equation}
For $\lambda=0$, the two late-time branches contribute equally, $\left|A_+/A_-\right|^2=1$, thereby reproducing the pure ghost result. When $\lambda\neq0$, this degeneracy is lifted, leading to an asymmetry between the two oscillatory boundary components. Its role is not to generate a real turning point for the heavy mode nor to induce a physical evanescent phase. Rather, it modifies the Whittaker connection coefficients, thereby biasing the relative weight of the two oscillatory branches appearing in the late-time expansion.

It is important to clarify the terminology. In what follows, we refer to $\lambda$ primarily as a branch-asymmetry parameter, or equivalently as an effective bias parameter, rather than as a genuine chemical potential. Its origin is kinematic: in the present setup, $\lambda=s_{\rm tilt}\beta/(4\gamma)$ is fixed by the sign and magnitude of the quadratic correction to the tilted-ghost dispersion relation. It therefore controls how the pure-ghost WKB branch selected in the far past is projected onto the two independent late-time Whittaker branches. This should be distinguished from standard chemical-potential mechanisms in cosmological-collider physics, where the Whittaker parameter is generated by the coupling of the rolling inflaton background to a conserved current, charge sector, or helicity structure of the heavy field. In particular, $\lambda$ is not associated with a conserved $U(1)$ charge, a helicity asymmetry, or an independent energy scale that sources heavy particles.

The appearance of $\nu/2$ in Eq.~\eqref{eq:TG_mode_normalized_compact} makes this distinction explicit. In the standard Whittaker realization of chemical-potential collider signals, the Whittaker variable is linear in conformal time, and the connection coefficients involve combinations of the form $\lambda\pm\mu$. In the tilted-ghost case, instead, the Whittaker variable is $z=i\gamma k^2\eta^2$, so the Whittaker index is $\nu/2=i\mu/2$, and the connection coefficients involve $\lambda\pm\mu/2$. This factor of $1/2$ does not change the physical clock frequency: after converting back from $z$ to $\eta$, the late-time branches remain $(-\eta)^{3/2\pm i\mu}$, and the squeezed-limit signal still oscillates with frequency $\mu$. It only reflects the non-relativistic origin of the Whittaker equation and the different projection of the early-time branch onto the late-time oscillatory modes.

\begin{figure}[ht]
\centering
{\includegraphics[width=10cm,height=6cm]{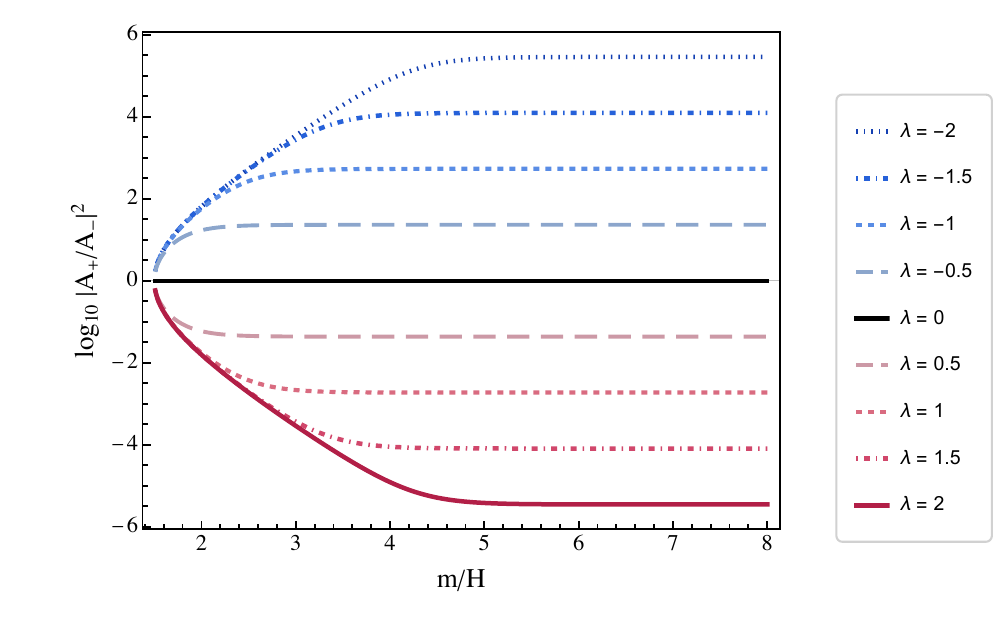}}
\caption{Branch asymmetry ratio for tilted ghost modes, shown through
\(\log_{10}|A_+/A_-|^2\), as a function of the exchanged mass. The pure ghost
case, \(\lambda=0\), gives equal weights for the two late-time oscillatory
branches and therefore appears as a constant line at zero. Nonzero values of
\(\lambda\) lift this degeneracy: one sign enhances the \(A_+\) branch relative
to \(A_-\), while the opposite sign suppresses it. This makes explicit the
chemical-potential-like role of \(\lambda\) in the Whittaker connection
coefficients. }
\label{asymm}
\end{figure}

This mechanism is closely analogous to the chemical-potential enhancement encountered in cosmological collider physics \cite{Qin_2023,Qin:2025xct,Bodas_2021}. This effect is illustrated in  Fig.~\ref{asymm}, where we plot $\log_{10}|A_+/A_-|^2$ as a function of the exchanged mass for different values of $\lambda$. The pure ghost case corresponds to the horizontal line at zero, while positive and negative values of $\lambda$ generate opposite asymmetries. As the mass increases, the curves approach an approximately constant offset, indicating that the chemical-potential-like bias persists deep into the heavy regime. Consequently, although the heavy mode remains oscillatory all the way to the boundary, the tilt leaves a measurable imprint through the relative enhancement of the two clock-signal branches.

The essential point is, therefore, that within the controlled regime defined by Eq.~\eqref{eq:heavy_no_turning_condition}, the parameter $\lambda\propto\beta/4\gamma$ does not signal the appearance of a real turning point in the heavy sector. Rather, it parametrizes a crossover in the relative weighting of the two oscillatory late-time branches. Accordingly, the notion of crossover for heavy tilted ghost modes should not be interpreted as the onset of a genuine evanescent phase, but instead as a redistribution of the Whittaker connection coefficients governing the asymptotic boundary solutions.

\section{Cosmological Collider}
\label{sec:collider}

The basic premise of the cosmological collider program is that interactions among fields during inflation are encoded in late-time cosmological correlators. In particular, particles propagating in the bulk can imprint characteristic signatures on boundary observables through their exchange contributions. 

From the perspective of perturbation theory, the relevant ingredients are the free mode functions, the propagators defined along the Schwinger--Keldysh contour \cite{Chen:2009zp,Chen_2012,Noumi_2013,Chen:2017ryl,Baumann_2018}, and the interaction Hamiltonian governing the time evolution of the quantum state. Since our aim is to investigate how bulk interactions are encoded in late-time observables, it is convenient to briefly review the \emph{in--in} formalism \cite{Weinberg_2005,calzeta,Chen_2010,Chen:2017ryl} in a form tailored to exchange diagrams and boundary correlators. 

In cosmology, the correlation functions are evaluated on a fixed late-time slice rather than in transition amplitudes between asymptotic in and out states. For this reason, the quantum state must be evolved both forward and backward in time along a closed contour, which naturally leads to the doubled field content of the \emph{in--in} construction. We begin by splitting the field into a homogeneous background and a fluctuation
\begin{equation}
\phi(\eta,\mathbf{x})=\phi_{0}(\eta)+\varphi(\eta,\mathbf{x}) \ .
\end{equation}

In this framework, the background $\phi_{0}(\eta)$ is considered a classical field, while the fluctuation $\varphi(\eta,\mathbf{x})$ is promoted to a quantum field. Expanding the action around the background determines the quadratic theory, from which the mode functions and propagators are constructed, together with the higher-order interaction vertices responsible for generating non-Gaussian observables. For the class of processes relevant to cosmological collider signals, it is useful to distinguish between fields that propagate exclusively in the bulk and fields whose fluctuations remain supported at the late-time boundary. Denoting by $\sigma$ a massive bulk field, the associated bulk-to-bulk propagators defined on the Schwinger--Keldysh contour take the form
\begin{align}
G_{++}(k,\eta_{1},\eta_{2})
&=
\sigma_{k}(\eta_{1})\sigma_{k}^{*}(\eta_{2})\Theta(\eta_{1}-\eta_{2})
+
\sigma_{k}^{*}(\eta_{1})\sigma_{k}(\eta_{2})\Theta(\eta_{2}-\eta_{1}) \, ,
\nonumber\\
G_{--}(k,\eta_{1},\eta_{2})
&=
\sigma_{k}(\eta_{1})\sigma_{k}^{*}(\eta_{2})\Theta(\eta_{2}-\eta_{1})
+
\sigma_{k}^{*}(\eta_{1})\sigma_{k}(\eta_{2})\Theta(\eta_{1}-\eta_{2}) \, ,
\label{eq:SKbulkbulk}
\\
G_{+-}(k,\eta_{1},\eta_{2})
&=
\sigma_{k}^{*}(\eta_{1})\sigma_{k}(\eta_{2}) \, ,
\nonumber\\
G_{-+}(k,\eta_{1},\eta_{2})
&=
\sigma_{k}(\eta_{1})\sigma_{k}^{*}(\eta_{2}) \, .
\nonumber
\end{align}
In the above expressions, $\Theta(\eta)$ denotes the Heaviside step function. The labels $\pm$ indicate whether the corresponding operator is inserted on the forward or backward branch of the time contour. Accordingly, $G_{++}$ is the time-ordered propagator, $G_{--}$ is the anti-time-ordered propagator, and $G_{+-}$ and $G_{-+}$ are the two Wightman functions. These satisfy the usual relations
\begin{equation}
G_{++}(k,\eta_{1},\eta_{2})=G_{--}^{*}(k,\eta_{1},\eta_{2}) \, ,
\qquad
G_{+-}(k,\eta_{1},\eta_{2})=G_{-+}^{*}(k,\eta_{1},\eta_{2}) \, .
\end{equation}
If $\varphi_k(\eta)$ describes a light external field whose correlators are evaluated at late times, we define the corresponding bulk-to-boundary propagators by taking one endpoint of the Schwinger-Keldysh propagator to the late time boundary $\eta_{0}\to 0$:
\begin{align}\label{eq:bulkboundary}
K_{+}(k,\eta)=\varphi_{k}(\eta_{0})\,\varphi_{k}^{*}(\eta) 
&&\, ,&&
K_{-}(k,\eta)=\varphi_{k}^{*}(\eta_{0})\,\varphi_{k}(\eta) 
\quad .
\end{align}
These kernels satisfy the relation $K_{+}(k,\eta)=K_{-}^{*}(k,\eta)$. They encode the propagation between a bulk interaction vertex and an operator insertion on the late-time boundary where the observable is evaluated. The remaining ingredient is the specification of the interaction Hamiltonian $H_I$, from which expectation values of operators evaluated at conformal time $\eta$ may be computed via the master \emph{in--in} formula
\begin{equation}
\langle \mathcal{O}(\eta)\rangle
=
\langle  \Omega |
\bar{T}
\exp\!\left(
i\int_{-\infty(1-i\epsilon)}^{\eta} d\eta'\, H_{I}(\eta')
\right)
\mathcal{O}_{I}(\eta)
T
\exp\!\left(
-i\int_{-\infty(1+i\epsilon)}^{\eta} d\eta'\, H_{I}(\eta')
\right)
| \Omega \rangle ,
\label{eq:ininmaster}
\end{equation}
where $T$ and $\bar{T}$ denote time-ordering and anti-time-ordering, respectively, and $|\Omega\rangle$ is the interacting vacuum selected by the standard $i\epsilon$ prescription. Expanding Eq.~\eqref{eq:ininmaster} perturbatively generates the exchange and contact contributions to cosmological correlators.  In the following sections, we specify the interactions of interest and apply this formalism to compute the corresponding late-time observables.

\subsection{Linear Mixing}
\label{sec:bispectrum_exchange}

To investigate the phenomenological consequences of the tilted-ghost spectator, we focus on its imprint on the inflationary bispectrum generated by single-particle exchange. The distinctive signature of the exchanged field is most clearly identified in the squeezed limit, where the momentum carried by the internal line becomes soft. We therefore consider a minimal set of mixing and cubic interactions between the external massless inflaton fluctuation $\varphi$ and the spectator field $\sigma$,
\begin{equation}\label{eq:int_L}
\delta\mathcal{L}_2 \supset 
\rho\,a^3\,\partial_{\eta}\varphi\,\sigma
\qquad , \qquad 
\delta\mathcal{L}_3 \supset g\,a^2\,(\partial_{\eta}\varphi)^2\,\sigma
\quad .
\end{equation}
where $a$ is the scale factor, and $\rho$ and $g$ are time-independent couplings. The coefficients $\rho$ and $g$ are treated as Wilson coefficients of the effective interaction basis, with their signs included in their definitions.\footnote{In cosmic time, the operators in Eq.~\eqref{eq:int_L} correspond schematically to $\rho\,  \dot\phi \, \sigma$ and $g \, \dot\phi^{2} \sigma$. For canonically normalized fields in four spacetime dimensions, $[\phi]=[\sigma]=1$, so $[\rho]=1$ and $[g]=-1$. Thus $g$ may be parameterized as $g=c_g/\Lambda_g$, while $\rho$ may arise, for instance, after expanding a higher-dimensional operator around the rolling background, giving $\rho\sim c_\rho\dot\phi_0/\Lambda_\rho$. The overall signs are convention-dependent and are absorbed into $\rho$ and $g$. The interaction scales $\Lambda_g$ and $\Lambda_\rho$ need not coincide with the ghost scale $M$ entering $\gamma=H/M$.}

These interactions generate the tree-level exchange diagram containing a single $\sigma$ propagator in the bulk and three external $\varphi$ legs evaluated at the late-time boundary. With the scale-factor structure in Eq.~\eqref{eq:int_L}, the cubic vertex involving two hard external derivatives contributes a factor $a^2(\eta_1)$, whereas the linear mixing vertex involving the soft external derivative contributes a factor $a^3(\eta_2)$, as used in the Schwinger--Keldysh exchange integral below. For the massless field $\varphi$, the bulk-to-boundary propagators are given by
\begin{equation}
K_\pm(k,\eta)
=
\frac{H^2}{2k^3}\,\bigl(1\mp i  k\eta\bigr)\,e^{\pm i k\eta}
\quad ,
\label{eq:K_massless}
\end{equation}

The exchanged tilted-ghost spectator is described by the bulk-to-bulk propagator $G_{ab}$, given by Eq~(\ref{eq:SKbulkbulk}), which we write in terms of Whittaker functions as
\begin{align}
\sigma_k(\eta_1)\sigma_k^*(\eta_2)
&=
\frac{H^2\,e^{-\pi\lambda}}{2\gamma k^2}\,
(\eta_1\eta_2)^{1/2}
\mathcal{W}_{-i\lambda,\, i\mu/2}\!\big(i\gamma k^2\eta_1^2\big)
\mathcal{W}_{+i\lambda,\,-i\mu/2}\!\big(-i\gamma k^2\eta_2^2\big)
\quad ,
\label{eq:G_TG}
\end{align}
where $\lambda\equiv s_{\rm tilt}\, {\beta}/{4\gamma}$. In \eqref{eq:G_TG} we have displayed the dependence on the tilted-ghost parameters through $\gamma$ and the signed dimensionless ratio $\lambda=s_{\rm tilt}\beta/(4\gamma)$, while the mass dependence enters through $\nu=i\mu$ (principal series) in the Whittaker indices. The exchange contribution to the three-point function takes the standard form
\cite{Pinol_2023,Lee_2016,Arkani-Hamed:2018kmz,Arkani-Hamed:2015bza}
\begin{align}
\big\langle
\varphi_{\mathbf{k}_1}
\varphi_{\mathbf{k}_2}
\varphi_{\mathbf{k}_3}
\big\rangle'_{\sigma}
=
-2\sum_{a,b=\pm}
\int_{-\infty}^{0}\! d\eta_1
&\int_{-\infty}^{0}\! d\eta_2\;
 a^2(\eta_1)a^3(\eta_2)\,
\partial_{\eta_1}K_a(k_1,\eta_1)\,
\partial_{\eta_1}K_a(k_2,\eta_1)
\nonumber\\
&\times
G_{ab}(s,\eta_1,\eta_2)\,
\partial_{\eta_2}K_b(k_3,\eta_2)
+2~\text{perm.}
\quad ,
\label{eq:inin_exchange_phi}
\end{align}
where $s\equiv|\mathbf{k}_1+\mathbf{k}_2|$ is the exchanged momentum and
$a,b=\pm$ are the Schwinger-Keldysh indices. Throughout, we keep the contour indices $a,b=\pm$ explicit to emphasize that the exchange contribution is obtained by summing over the in-in polarities. In this expression, and in Eq.~\eqref{integregr} below, we suppress the overall product of couplings and numerical symmetry factors; these are absorbed into the overall normalization of the shape function in Section~\ref{sec:phenomenology}. The three-point function reads
\begin{equation}\label{integregr}
\langle
\varphi_{\mathbf k_1}
\varphi_{\mathbf k_2}
\varphi_{\mathbf k_3}
\rangle^{\prime}_{\sigma}
=
\frac{H}{4k_1k_2k_3}
\sum_{a,b=\pm1}ab
\int d\eta_1\,
\frac{d\eta_2}{(-\eta_2)^2}\,
e^{iak_{12}\eta_1+ibk_3\eta_2}
G_{ab}(s,\eta_1,\eta_2)
+2\,\text{perm.}
\end{equation}
It is useful to make the three exchange channels explicit. Here, we use the notation $k_{ij}\equiv k_i+k_j$ and $s_{ij}\equiv |\mathbf k_i+\mathbf k_j|$. Using momentum conservation, one has $s_{12}=k_3$, $s_{23}=k_1$, and $s_{31}=k_2$. Therefore, the full correlator can be written as
\begin{equation}
\langle
\varphi_{\mathbf k_1}
\varphi_{\mathbf k_2}
\varphi_{\mathbf k_3}
\rangle'_{\sigma}
=
\mathcal C_{12|3}
+\mathcal C_{23|1}
+\mathcal C_{31|2},
\label{corrdecomp}
\end{equation}
where each term corresponds to a distinct exchange channel. In the following sections, we analyze the contribution of each channel to the soft and hard components of the signal.

\subsection{Soft/collider contribution}
\label{subsec:soft_collider_main}

We now separate the part of the exchange diagram that carries the cosmological collider signal.  In the squeezed configuration
\begin{align}
 k_3\ll k_1\sim k_2,
\end{align}
the channel \(\mathcal C_{12|3}\) is special because the exchanged momentum is
\begin{align}
 s_{12}=|\mathbf k_1+\mathbf k_2|=k_3\to0 .
\end{align}
The other two exchange channels have internal momenta \(s_{23}=k_1\) and
\(s_{31}=k_2\), and therefore remain hard in the squeezed limit. For the soft channel, we define
\begin{align}
K\equiv k_1+k_2
\quad ,
\qquad
\Delta_\pm\equiv \frac32\pm i\mu 
\quad .
\end{align}
The channel contribution can be written as
\begin{align}
\mathcal C_{12|3}
=
\frac{H}{4k_1k_2k_3}
\sum_{a,b=\pm1}ab\,
\mathcal I^{ab}_{12|3}
\quad ,
\end{align}
where
\begin{align}
\mathcal I^{ab}_{12|3}
=
\int_{-\infty}^{0}d\eta_1
\int_{-\infty}^{0}\frac{d\eta_2}{(-\eta_2)^2}\,
e^{iaK\eta_1+ibk_3\eta_2}
G_{ab}(k_3,\eta_1,\eta_2)
\quad .
\label{eq:soft_channel_integral_main}
\end{align}

The complete Schwinger-Keldysh bookkeeping is presented in Appendix~\ref{app:sk_exchange}. For the present discussion, the only relevant point is that, in the squeezed limit $s_{12}=k_3\to0$, the internal propagator may be approximated by its late-time expression. We parameterize this form as
\begin{align}
\mathcal G_{\rm LT}(k_3;\eta_1,\eta_2)
&=
\mathcal A_- k_3^{-2i\mu}
(-\eta_1)^{\Delta_-}(-\eta_2)^{\Delta_-}
+
\mathcal B_{-+}
(-\eta_1)^{\Delta_-}(-\eta_2)^{\Delta_+}
\nonumber\\
&\quad+
\mathcal A_+ k_3^{2i\mu}
(-\eta_1)^{\Delta_+}(-\eta_2)^{\Delta_+}
+
\mathcal B_{+-}
(-\eta_1)^{\Delta_+}(-\eta_2)^{\Delta_-}
\quad .
\label{eq:GLT_main}
\end{align}

Here $\mathcal A_+ = \mathcal A_-^\ast$ for real $\lambda$ and $\mu$, whereas $\mathcal B_{-+}$ and $\mathcal B_{+-}$ are both real but generically unequal. The relation $\mathcal B_{-+}\neq\mathcal B_{+-}$ is crucial, since the mixed branches are not symmetric under the exchange $\eta_1\leftrightarrow\eta_2$. This asymmetry constitutes the key bookkeeping ingredient responsible for fixing the powers of $k_3$ in the squeezed limit. The elementary Schwinger--Keldysh integrals reduce to monomial integrals of the form $(-\eta_1)^p(-\eta_2)^q$. As shown in Appendix~\ref{app:sk_exchange}, both the factorized contributions and the non-analytic parts of the ordered components produce the same momentum scaling, $K^{-1-p}k_3^{,1-q}$, up to contour-dependent phases. Applying this scaling rule to the four late-time branches appearing in Eq.~\eqref{eq:GLT_main}, one obtains the pair of complex-conjugate non-analytic structures $K^{-5/2+i\mu}k_3^{-1/2-i\mu}$ and $K^{-5/2-i\mu}k_3^{-1/2+i\mu}$. These terms correspond to the logarithmic clock branches.
The explicit evaluation of each contribution combines to yield
\begin{align}
\mathcal I_{\rm SK}
&\simeq
2\,
\Gamma\!\left(\frac52-i\mu\right)
K^{-5/2+i\mu}k_3^{-1/2-i\mu}
\Bigg[
\bigl(1-i\sinh\pi\mu\bigr)
\mathcal A_-\Gamma\!\left(\frac12-i\mu\right)
+
\mathcal B^{\rm SK}_{-}
\Gamma\!\left(\frac12+i\mu\right)
\Bigg]
\nonumber\\
&+2\,
\Gamma\!\left(\frac52+i\mu\right)
K^{-5/2-i\mu}k_3^{-1/2+i\mu}
\Bigg[
\bigl(1+i\sinh\pi\mu\bigr)
\mathcal A_+\Gamma\!\left(\frac12+i\mu\right)
+
\mathcal B^{\rm SK}_{+}
\Gamma\!\left(\frac12-i\mu\right)
\Bigg],
\label{eq:ISK_main_result}
\end{align}
where
\begin{align}
\mathcal B^{\rm SK}_{\pm}
&=
\frac{H^2}{4\mu}\,
\csch(\pi\mu)
\left[
1+e^{-2\pi\lambda}\cosh(\pi\mu)
\right]
\pm
i\,\frac{H^2}{4\mu}.
\label{eq:BSK_main}
\end{align}
For real $\lambda$ and $\mu$, $\mathcal B^{\rm SK}_{+}=(\mathcal B^{\rm SK}_{-})^*$. The details leading to Eq.~\eqref{eq:ISK_main_result} are collected in
Appendix~\ref{subsec:soft_whittaker_corrected}. 
Notice that Eq.~\eqref{eq:ISK_main_result} contains two distinct classes of soft contributions with different physical origins. The terms proportional to $\mathcal A_\pm$ correspond to the genuine Whittaker-branch contributions: their non-analytic behavior is already encoded in the small-momentum expansion of the propagator through the factors $k_3^{\pm2i\mu}$. By contrast, the terms proportional to $\mathcal B^{\rm SK}_\pm$ represent mixed soft contributions. These originate from the unequal-power branches of the late-time propagator and acquire the same clock-like scaling only after performing the Schwinger--Keldysh time integrals. 

In the discussion below, the term ``non-analytic'' refers specifically to the Whittaker-branch contribution controlled by $A_\pm$. Although the mixed soft terms are retained in the complete shape coefficients presented below, they should not be identified either with the genuine Whittaker-branch non-analytic contribution or with the analytic hard-channel background.

The non-analytic contribution associated with the oscillatory Whittaker branches may be expressed in a compact form. For the tilted-ghost mode considered in this work, the resulting expression is
\begin{align}
\mathcal I_{{\rm TG},\, {\rm non\mbox{-}an}}
&\simeq
\frac{H^2}{e^{\pi\lambda}k_3^3}
\left(\frac{k_3}{K}\right)^{5/2}
\left(\frac{\gamma k_3}{K}\right)^{-i\mu}
\frac{
\bigl[1-i\sinh(\pi\mu)\bigr]
\Gamma\!\left(\frac12-i\mu\right)
\Gamma\!\left(\frac52-i\mu\right)
\Gamma(i\mu)^2
}{
\Gamma\!\left(\frac12-i\lambda+\frac{i\mu}{2}\right)
\Gamma\!\left(\frac12+i\lambda+\frac{i\mu}{2}\right)
}
+
{\rm c.c.}
\label{eq:TG_whittaker_nonlocal_main}
\end{align}
The positive-tilt result is obtained through the replacement $\lambda\to-\lambda$. In the squeezed configuration, one may further approximate $K\simeq2k_1$. Taking the pure-ghost limit $\lambda\to0$ yields
\begin{align}
&\mathcal I_{{\rm PG},\,  {\rm non\mbox{-}an}}
\equiv
\lim_{\lambda\to0}
\mathcal I_{{\rm TG},\, {\rm non\mbox{-}an}}
\label{eq:TG_pure_ghost_limit_main}\\
& \qquad \simeq
\frac{H^2}{4\pi k_3^3}
\left(\frac{k_3}{K}\right)^{5/2}
\left(\frac{\gamma k_3}{4K}\right)^{-i\mu}
\bigl[1-i\sinh(\pi\mu)\bigr]
\Gamma\!\left(\frac12-i\mu\right)
\Gamma\!\left(\frac52-i\mu\right)
\Gamma\!\left(\frac{i\mu}{2}\right)^2
+{\rm c.c.}
\nonumber
\end{align}
where we used the duplication formula to rewrite the result in the Hankel-like pure-ghost normalization. The overall normalization coincides with that adopted in Eq.~\eqref{eq:TG_whittaker_nonlocal_main}. Changing the normalization of the bulk mode merely rescales the overall prefactor, leaving the clock exponent unchanged. 

It is also instructive to present the analogous expression for the standard Whittaker mode (see Appendix~\ref{subsec:soft_whittaker_corrected}), commonly employed in boostless or chemical-potential cosmological-collider templates. Using the standard Whittaker normalization parameterized by $c_b$, one obtains
\begin{align}
\mathcal I_{{\rm W},\, {\rm non\mbox{-}an}}
&\simeq
\frac{2H^2e^{-\pi\lambda}}{k_3^3}
\left(\frac{k_3}{K}\right)^{5/2}
\left(4c_b^2\frac{k_3}{K}\right)^{-i\mu}
\nonumber\\
&\quad\times
\frac{
\bigl[1-i\sinh(\pi\mu)\bigr]
\Gamma\!\left(\frac12-i\mu\right)
\Gamma\!\left(\frac52-i\mu\right)
\Gamma(2i\mu)^2
}{
\Gamma\!\left(\frac12-i\lambda+i\mu\right)
\Gamma\!\left(\frac12+i\lambda+i\mu\right)
}
+
{\rm c.c.}
\label{eq:standard_whittaker_nonlocal_main}
\end{align}
In the corresponding Hankel limit, obtained by taking $\lambda \to 0$, this expression reduces to
\begin{align}
\mathcal I_{{\rm H},\, {\rm non\mbox{-}an}}
&\simeq
\frac{H^2}{2\pi k_3^3}
\left(\frac{k_3}{K}\right)^{5/2}
\left(\frac{c_b^2 k_3}{4K}\right)^{-i\mu}
\nonumber\\
&\quad\times
\bigl[1-i\sinh(\pi\mu)\bigr]
\Gamma\!\left(\frac12-i\mu\right)
\Gamma\!\left(\frac52-i\mu\right)
\Gamma(i\mu)^2
+
{\rm c.c.}
\label{eq:standard_hankel_limit_main}
\end{align}
Equations~\eqref{eq:TG_whittaker_nonlocal_main}-\eqref{eq:standard_hankel_limit_main} make it clear that the tilted-ghost Whittaker result cannot be obtained from the standard Whittaker expression through a simple redefinition of parameters. Although both expressions exhibit the same logarithmic clock structure, they differ in both their Whittaker indices and their ultraviolet normalizations. In the tilted-ghost case, the Whittaker parameter is determined by the ratio $\lambda=\beta/4\gamma$, and the pure-ghost limit is governed by the underlying quartic dispersion relation. By contrast, in the standard Whittaker case, the Hankel limit corresponds to the relativistic, boostless reference solution encoded by the parameter $c_b$.

\subsection{Hard/background contributions}
\label{subsec:hard_background_main}

We now analyze the contributions from the two remaining permutations, $\mathcal C_{23|1}$ and $\mathcal C_{31|2}$. In the squeezed limit, these correspond to hard exchange channels, since the exchanged momenta are $s_{23}=k_1$ and $s_{31}=k_2$, respectively. Consequently, the late-time small-$s$ expansion of the internal propagator is not applicable in these channels, and the full Whittaker bulk-to-bulk propagator must be retained. For example,
\begin{align}
\mathcal I^{ab}_{23|1}
=
\int_{-\infty}^{0}d\eta_1
\int_{-\infty}^{0}\frac{d\eta_2}{(-\eta_2)^2}\,
e^{ia(k_2+k_3)\eta_1+ibk_1\eta_2}
G_{ab}(k_1,\eta_1,\eta_2),
\end{align}
and similarly for $\mathcal C_{31|2}$ with $k_1\leftrightarrow k_2$.  After the rescaling $x=-k_1\eta_1$, $y=-k_1 \eta_2$, and $r=(k_2+k_3)/k_1$, the first hard channel becomes
\begin{align}
\mathcal C_{23|1}
=
\frac{H}{4k_1k_2k_3}
\sum_{a,b=\pm}
\int_0^\infty dx
\int_0^\infty\frac{dy}{y^2}\,
e^{-iarx-iby}
G_{ab}(x,y;\lambda,\gamma)
\quad .
\end{align}
Since $r=1+\mathcal O(k_3/k_1)$, the hard-channel contribution admits an analytic expansion in the squeezed ratio, which may be written as
\begin{align}
\left\langle
\varphi_{\mathbf k_1}
\varphi_{\mathbf k_2}
\varphi_{\mathbf k_3}
\right\rangle'_{\rm hard}
\equiv
\mathcal C_{23|1}+\mathcal C_{31|2}
\simeq
2\,e^{-\pi\lambda}
\sum_{n=0}^{\infty}
C_n(\mu,\lambda,\gamma)
\left(\frac{k_3}{k_1}\right)^n 
\quad .
\label{eq:hard_series_main}
\end{align}
The explicit Whittaker representation of the coefficients $C_n(\mu,\lambda,\gamma)$ is given in Appendix~\ref{app:hard_channels}. In contrast to the soft channel, Eq.~\eqref{eq:hard_series_main} is analytic in the squeezed ratio $k_3/k_1$. It therefore contributes to the analytic background of the squeezed bispectrum, whereas the non-analytic clock signal is carried by $\mathcal C_{12|3}$. The factor $e^{-\pi\lambda}$ in Eq.~\eqref{eq:hard_series_main} is the universal normalization inherited from the tilted Wightman propagator. Although this factor can be pulled outside the Schwinger--Keldysh sums and time integrals, the tilt dependence is not removed. The remaining dependence on $\lambda$ and $\gamma$ persists through the Whittaker indices and through the arguments $i\gamma x^2$ and $-i\gamma y^2$. Thus, the hard-channel terms retain information about the tilted dispersion through the internal propagator, but they do not alter the external squeezed expansion parameter, which remains $k_3/k_1$.

\section{Phenomenology}
\label{sec:phenomenology}

To connect the exchange correlator computed for the inflaton fluctuation $\varphi$ with observable quantities, it is necessary to express the result in terms of the comoving curvature perturbation $\zeta$, which remains conserved on super-horizon scales and directly determines the primordial power spectrum and higher-order correlation functions. In the present conventions, $\zeta$ is related to the inflaton fluctuation by
\begin{equation}
\zeta=\frac{\varphi}{\sqrt{2\epsilon} \ M_{\rm Pl}},
\label{eq:zeta_phi}
\end{equation}
where $\epsilon$ denotes the slow-roll parameter and $M_{\rm Pl}$ is the reduced Planck mass. We assume standard slow-roll inflationary dynamics; hence, the power spectrum reads
\begin{equation}
P_\zeta=
\frac{H^2}{8\pi^2\epsilon M_{\rm Pl}^2}
\quad .
\end{equation}
These relations provide the dictionary between the exchange correlators computed for $\varphi$ and the corresponding bispectrum and shape function for $\zeta$. The bispectrum of the curvature perturbation is defined through the three-point correlation function in momentum space as
\begin{equation}
 B_\zeta(k_1,k_2,k_3)
 \equiv
 \big\langle
 \zeta_{\mathbf{k}_1}
 \zeta_{\mathbf{k}_2}
 \zeta_{\mathbf{k}_3}
 \big\rangle'
 \quad .
\label{eq:bispectrum_def}
\end{equation}
It is convenient to introduce a dimensionless shape function $S$ via
\begin{equation} \label{eq:shape_def}
\big\langle
\zeta_{\mathbf{k}_1}
\zeta_{\mathbf{k}_2}
\zeta_{\mathbf{k}_3}
\big\rangle'
=
(2\pi)^4
\frac{P_\zeta^2}{(k_1k_2k_3)^2}
S(k_1,k_2,k_3)
\quad 
.
\end{equation}
Our goal is to isolate the non-analytic momentum dependence in the squeezed limit since this component provides the cleanest diagnostic of particle-exchange physics. The simplified shapes introduced below should therefore be regarded as analytic templates for the propagation-induced signal. Data-oriented analyzes require full-shape treatments or survey-specific implementations, as developed in recent CMB, trispectrum, and large-scale structure searches~\cite{Sohn:2024planck,Suman:2025significance,Suman:2025cmbII,Cabass:2024boss,Philcox:2025cmbtrispectrumI,Philcox:2025cmbtrispectrumII,Green:2026extending,Kumar:2026beyond,Philcox:2025cmbtrispectrumIII,Goldstein:2024massiveish,Goldstein:2025wiggly,Kumar:2026scalars}.

In cosmological collider physics, the squeezed limit corresponds to the regime of primordial correlators in which one momentum is much smaller than the others, namely $k_L \equiv k_3 \ll k_1 \simeq k_2 \equiv k_S$, such that $k_L \ll k_S$. This limit is particularly important because it isolates the contribution from particle exchange during inflation, making the characteristic non-analytic momentum dependence manifest. In this regime, the internal exchanged momentum satisfies $s \simeq k_L$, so that the long-wavelength mode directly probes the propagation of the intermediate particle. Consequently, the correlator shape admits a schematic factorization form
\begin{equation}
S_{\rm sq}(\lambda,\mu,\gamma)=\mathcal{N}\,F(\lambda,\mu,\gamma)
\quad ,
\label{eq:Sq_factorization}
\end{equation}
where $\mathcal{N}$ collects the overall slow-roll, coupling, and normalization factors, including the interaction coefficients $\rho$ and $g$ appearing in \eqref{eq:int_L}, while $F$ captures the non-analytic momentum dependence that remains in the soft limit and contains the characteristic particle-exchange signal. Note that the hard/background contribution remains analytic in \(k_3/k_1\), as in Eq.~(\ref{eq:hard_series_main}). But its overall amplitude still inherits the \(\lambda\)-dependent Whittaker normalization, exhibiting the same sign-dependent pattern that appears in the soft contribution. The plots below focus on the non-analytic soft channel.

\begin{figure}[t]
\hfill
{\includegraphics[width=7cm]{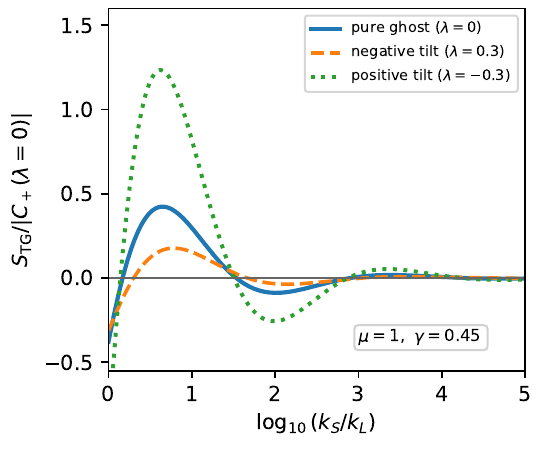}}
\hfill
{\includegraphics[width=7cm]{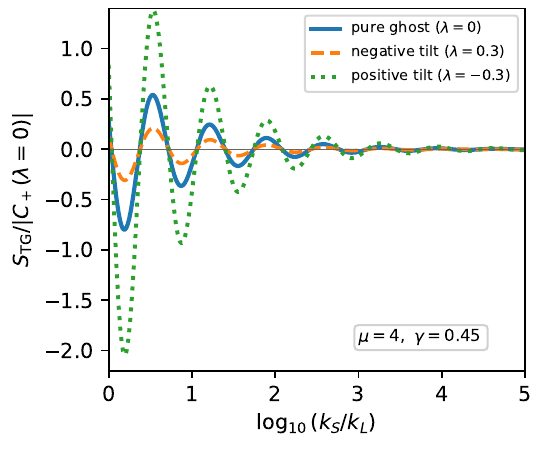}}
\hfill
\caption{Oscillatory soft contribution to the bispectrum for a modest signed tilt, compared with the pure-ghost exchange. The vertical axis is normalized by the pure-ghost amplitude, $|C_+(\lambda=0)|$, at the same values of $\mu$ and $\gamma$. This normalization removes the common ultraviolet pure-ghost normalization and isolates the relative effect of the signed tilt. In the convention of Eq.~\eqref{eq:TG_signal_real}, $\lambda<0$ corresponds to the positive-tilt branch and enhances the signal, while $\lambda>0$ corresponds to the negative-tilt branch and suppresses it. Left: $\mu=1$. Right: $\mu=4$.}
\label{bISPEC}
\end{figure}

We now express the non-analytic signal associated with the late-time result in the squeezed limit in the standard cosmological-collider form for the tilted-ghost scenario studied in the present work. In this regime, the momentum-dependent contributions appearing in $\mathcal I_{\rm SK}$ simplify considerably and reduce to the characteristic soft-momentum scaling structure
\begin{align}
k_1^2 k_3\,K^{-5/2+i\mu}k_3^{-1/2-i\mu}
&=
2^{-5/2}
\left(\frac{k_3}{k_1}\right)^{1/2}
\left(\frac{k_3}{2k_1}\right)^{-i\mu}
\quad ,
\\
k_1^2 k_3\,K^{-5/2-i\mu}k_3^{-1/2+i\mu}
&=
2^{-5/2}
\left(\frac{k_3}{k_1}\right)^{1/2}
\left(\frac{k_3}{2k_1}\right)^{i\mu}
\quad .
\end{align}
For the tilted-ghost mode functions, the non-analytic coefficients acquire an additional phase factor $\gamma^{\pm i\mu}$. By contrast, the SK-mixed contribution does not originally contain this phase. Therefore, when the full signal is rewritten in terms of the combination $\gamma k_3/k_1$, the mixed term acquires a compensating factor $\gamma^{\mp i\mu}$ in its coefficient. With this rearrangement, the signal can be written as
\begin{equation}
S_{\rm TG}
=
\frac12
\left(\frac{k_3}{k_1}\right)^{1/2}
\left[
\mathcal C_-^{\rm TG}
\left(
\gamma\frac{k_3}{k_1}
\right)^{-i\mu}
+
\mathcal C_+^{\rm TG}
\left(
\gamma\frac{k_3}{k_1}
\right)^{i\mu}
\right].
\label{eq:TG_signal_complex}
\end{equation}
where the two coefficients are
\begin{align}
\mathcal C_+^{\rm TG}
&=
2^{-7/2-i\mu}
\left(\frac12+i\mu\right)
\left(\frac32+i\mu\right)
\Bigg[
e^{-\pi\lambda}
\left(1+i\sinh\pi\mu\right)
\frac{
\Gamma(-i\mu)^2
\Gamma\!\left(\frac12+i\mu\right)^2
}{
\Gamma\!\left(\frac12-i\lambda-\frac{i\mu}{2}\right)
\Gamma\!\left(\frac12+i\lambda-\frac{i\mu}{2}\right)
}
\nonumber\\
&\qquad\qquad
+
\frac{\pi}{2\mu}\,
\gamma^{-i\mu}
\left\{
\frac{
1+e^{-2\pi\lambda}\cosh(\pi\mu)
}{
\sinh(\pi\mu)\cosh(\pi\mu)
}
+
i\,\operatorname{sech}(\pi\mu)
\right\}
\Bigg]
\label{eq:Cplus_TG_final}\\
\mathcal C_-^{\rm TG}
&=
2^{-7/2+i\mu}
\left(\frac12-i\mu\right)
\left(\frac32-i\mu\right)
\Bigg[
e^{-\pi\lambda}
\left(1-i\sinh\pi\mu\right)
\frac{
\Gamma(i\mu)^2
\Gamma\!\left(\frac12-i\mu\right)^2
}{
\Gamma\!\left(\frac12-i\lambda+\frac{i\mu}{2}\right)
\Gamma\!\left(\frac12+i\lambda+\frac{i\mu}{2}\right)
}
\nonumber\\
&\qquad\qquad
+
\frac{\pi}{2\mu}\,
\gamma^{i\mu}
\left\{
\frac{
1+e^{-2\pi\lambda}\cosh(\pi\mu)
}{
\sinh(\pi\mu)\cosh(\pi\mu)
}
-
i\,\operatorname{sech}(\pi\mu)
\right\}
\Bigg]
\label{eq:Cminus_TG_final}
\end{align}
For real $\lambda$, $\mu$, and $\gamma$, these coefficients satisfy $\mathcal C_-^{\rm TG}
=\left(\mathcal C_+^{\rm TG}\right)^\ast$. Thus, the signal may be written in the real cosmological-collider form
\begin{equation}
S_{\rm TG}
=
\left(\frac{k_3}{k_1}\right)^{1/2}
\left|\mathcal C_+^{\rm TG}\right|
\cos\left[
\mu\log\left(
\gamma\frac{k_3}{k_1}
\right)
+
\arg \mathcal C_+^{\rm TG}
\right].
\label{eq:TG_signal_real}
\end{equation}
The above template encodes the cosmological-collider contribution generated by tilted-ghost modes. These modes inherit the pure-ghost normalization in the far-ultraviolet regime, and therefore differ from the standard massive de Sitter or chemical-potential Whittaker modes, already at the level of the vacuum prescription. The signed parameter $\lambda = s_{\rm tilt}\, {\beta}/{4\gamma}$ controls how the pure-ghost positive-frequency branch is projected onto the two late-time principal-series branches. For $|\lambda|\ll1$, the tilt acts as a perturbative deformation of the pure-ghost signal. By contrast, when $|\lambda|\sim1$, the connection coefficients can be significantly reorganized, leading to either an enhancement or a suppression of the signal depending on the sign of $\lambda$. This sign dependence is clearly visible in the squeezed signal shown in Fig.~\ref{bISPEC}, where we consider a modest signed tilt. The signal is normalized by the pure-ghost amplitude, $|C_+(\lambda=0)|$, at fixed $\mu$ and $\gamma$. This choice factors out the common normalization inherited from the ultraviolet $k^4$ branch and isolates the relative effect of the signed tilt on the envelope and phase of the clock signal.
The branch with $\lambda<0$, corresponding to a positive tilt in the dispersion relation, enhances the oscillatory soft contribution relative to the pure-ghost result, whereas the branch with $\lambda>0$, corresponding to a negative tilt, suppresses it. Equivalently, in terms of the exact coefficients, reversing the sign of $\lambda$ modifies the Boltzmann-like weights associated with the late-time Whittaker branches. This hierarchy remains qualitatively stable as the mass parameter is varied: the same pattern is observed for both $\mu=1$ and $\mu=4$, although the heavier case exhibits more rapid oscillations in $\log(k_S/k_L)$.

\begin{figure}[t]
\hfill
{\includegraphics[width=7cm]{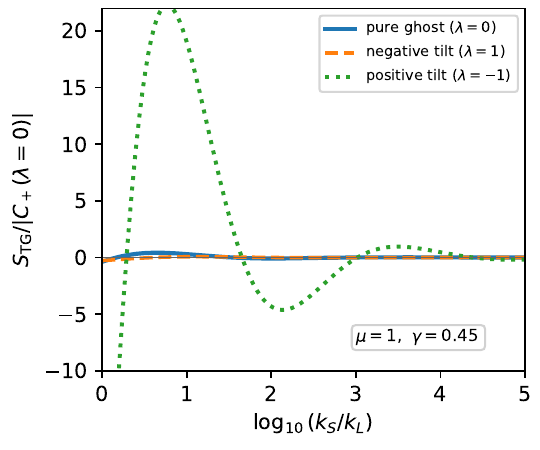}}
\hfill
{\includegraphics[width=7cm]{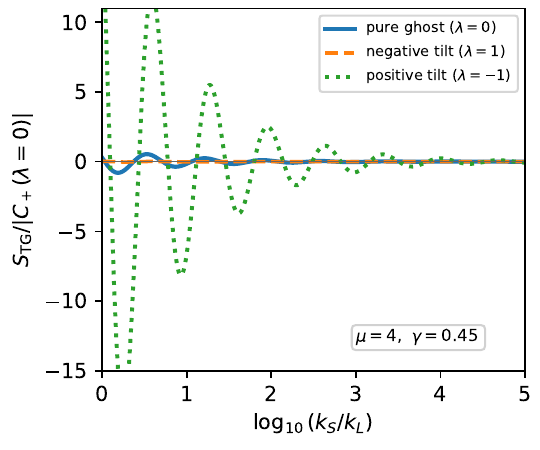}}
\hfill
\caption{Oscillatory soft contribution for an order-one signed tilt, normalized to the pure-ghost amplitude, $|C_+(\lambda=0)|$, at fixed $\mu$ and $\gamma$. This normalization removes the common pure-ghost ultraviolet normalization and makes the relative $\lambda$-dependent enhancement or suppression manifest. The plot should be read as an edge-of-regime diagnostic rather than as a parametrically small deformation. The enhancement of the $\lambda<0$ branch and the suppression of the $\lambda>0$ branch are now much more pronounced, reflecting the branch-asymmetry factors in the tilted-ghost Whittaker coefficients. Left: $\mu=1$. Right: $\mu=4$.}
\label{bISPEC2}
\end{figure}

For order-one values of the signed tilt, near the boundary of the controlled regime, the same mechanism becomes significantly more pronounced, as illustrated in Fig.~\ref{bISPEC2}. The cosmological clock signal continues to oscillate with frequency $\mu$. Instead, the effect originates from the $\lambda$-dependent Whittaker connection coefficients, which determine the amplitude envelope and phase of the late-time oscillatory signal. This also clarifies why the enhancement is not expected to persist generically outside the controlled tilted-ghost regime: once the negative-sign quadratic contribution becomes too large, the system can no longer be regarded as a perturbative deformation of the pure-ghost spectator sector.

Although the initial condition is fixed in the far-ultraviolet scale by the dominance of the $k^4$ term in the dispersion relation, the subsequent evolution is strongly influenced by the $k^2$ contribution. In the positive-tilt branch, this term enters with the opposite sign and may induce a transient instability. For this reason, it cannot be allowed to dominate over the quartic ghost contribution. This provides a physical motivation for restricting the parameter space to $|\lambda|=\beta/(4\gamma)\lesssim 1$, or, more conservatively, to the genuine small-tilt regime $|\lambda|=\beta/(4\gamma)\ll1$. Beyond this regime, the enhancement can no longer be interpreted as a controlled perturbative deformation of the pure-ghost spectator signal.

This also clarifies the status of possible backreaction effects. The transient instability of the positive-tilt branch is a feature of the light-field regime, where the wrong-sign quadratic term can generate a finite interval of growth before freeze-out. Such a regime is compatible with the spectator interpretation only if the amplified fluctuations remain energetically subdominant, so that $\rho_\sigma\ll 3M_{\rm Pl}^2H^2$, and if the tilt is kept within the controlled EFT domain. By contrast, for the principal-series modes relevant to the cosmological-collider signal, the positive contribution $(\mu^2+1/4)/\eta^2$ prevents a real evanescent interval as long as Eq.~\eqref{eq:heavy_no_turning_condition} is satisfied. Therefore, in the regime used for the heavy-particle exchange calculation, the tilt reorganizes the Whittaker connection coefficients rather than producing a tachyonic growth phase. Parameter choices outside this regime would require a dedicated analysis of particle production and backreaction, and are not used as conservative EFT benchmarks in this work.

\begin{figure}[t]
\centering
{\includegraphics[width=10cm]{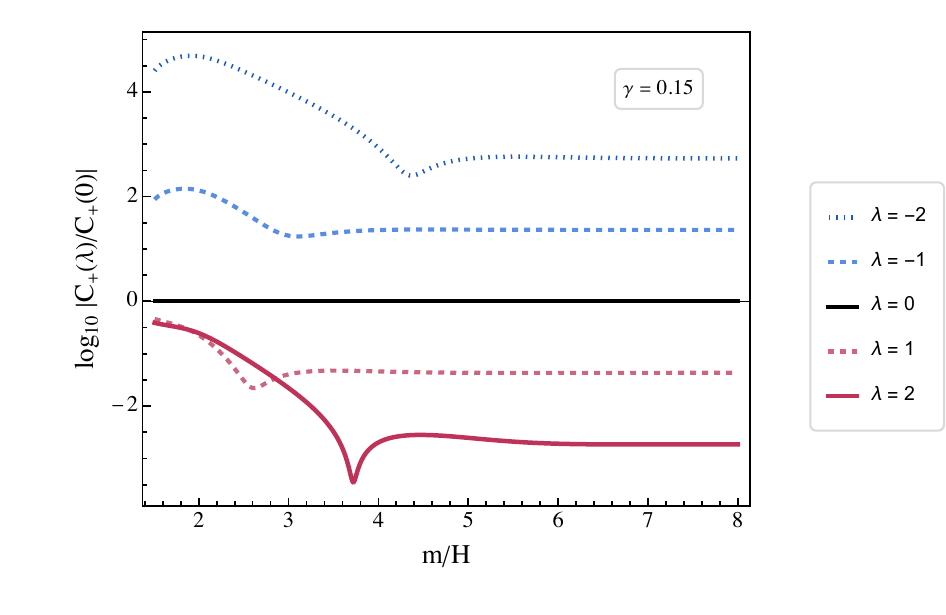}}
\caption{Relative envelope of the tilted-ghost clock signal, $\log_{10}|C_+(\lambda)/C_+(0)|$, as a function of the exchanged mass. The curves with $|\lambda|>1$ are included only to display the mathematical trend of the Whittaker connection coefficients and should not be interpreted as conservative EFT benchmarks. In particular, they can extend beyond the controlled small-tilt regime associated with Eq.~\eqref{eq:heavy_no_turning_condition}, namely $2|\lambda|<\sqrt{\mu^2+1/4}$. The signed tilt produces an asymmetric response: the $\lambda<0$ branch is enhanced, whereas the $\lambda>0$ branch is suppressed.}
\label{bISPEC3}
\end{figure}

The same behavior is summarized in the relative-amplitude plot shown in Fig.~\ref{bISPEC3}. Curves with $|\lambda|>1$ should be regarded primarily as diagnostics of the Whittaker connection coefficients rather than as conservative EFT benchmarks. The key point is that the amplitude hierarchy is largely determined by the choice of the ultraviolet branch. In absolute normalization, tilted-ghost and standard Whittaker templates may differ by large factors because they are associated with different early-time vacuum prescriptions: the tilted-ghost mode is matched to the pure-ghost WKB branch, whereas the standard Whittaker mode is anchored to a relativistic Bunch-Davies-like branch. Once the signal is normalized by its corresponding $\lambda=0$ amplitude, the large overall normalization difference is removed, and the remaining observable information is encoded primarily in the $\lambda$-dependent envelope and phase of the oscillatory signal.

The flattening at large mass can be understood directly from the large-$\mu$ asymptotics of the Whittaker connection coefficients, see Appendix \ref{app:large-mu-whittaker}. Using Stirling's approximation for the Gamma functions, and keeping fixed $\lambda$ with $\mu\gg |\lambda|$, the dominant Whittaker-branch contribution behaves as
\begin{equation}
\left|
\frac{C_+(\lambda,\mu,\gamma)}
{C_+(0,\mu,\gamma)}
\right|
\simeq
e^{-\pi\lambda}
\left[
1+\mathcal O\left(\frac{1}{\mu}\right)
+
\mathcal O\left(e^{-\pi\mu/2}\right)
\right].
\end{equation}
Therefore
\begin{equation}
\log_{10}
\left|
\frac{C_+(\lambda,\mu,\gamma)}
{C_+(0,\mu,\gamma)}
\right|
\longrightarrow
-\frac{\pi\lambda}{\ln 10},
\qquad
\mu\rightarrow\infty .
\end{equation}
This explains why each curve in Fig.~\ref{bISPEC3}approaches approximately a constant plateau at large $m/H$: after normalization by the $\lambda=0$ amplitude, the common large-$\mu$ suppression cancels, while the signed branch-asymmetry factor remains.

\begin{figure}[t]
\hfill
{\includegraphics[width=7cm]{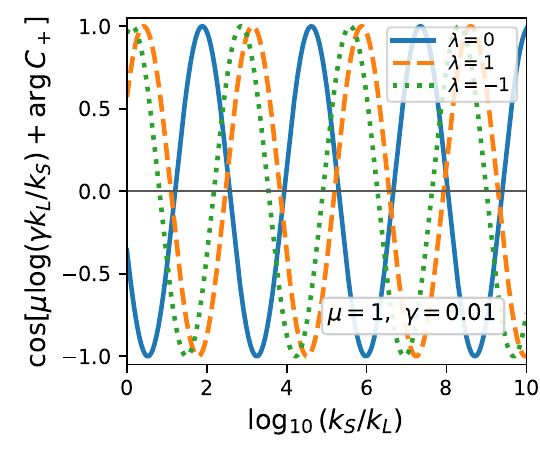}}
\hfill
{\includegraphics[width=7cm]{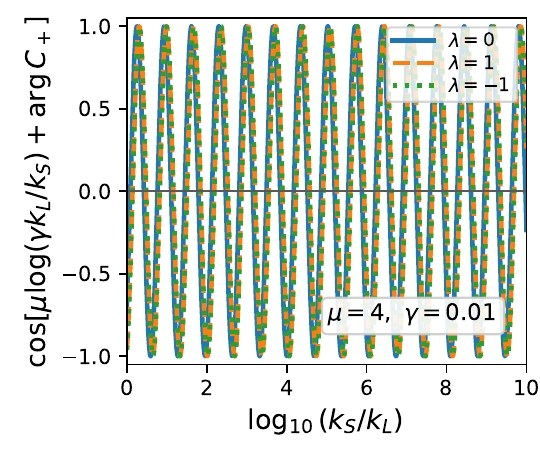}}
\hfill
\caption{Phase-only comparison of the oscillatory clock signal for the pure-ghost branch and representative positive- and negative-tilt deformations. The plotted quantity isolates the cosine in Eq.~\eqref{eq:TG_signal_real}, showing how the tilt changes \(\arg C_+\) and therefore displaces the signal in \(\log\kappa\)-space while leaving the clock frequency \(\mu\) fixed. The effect is easier to resolve for the lighter field, \(\mu=1\), and becomes visually less pronounced for \(\mu=4\), where the clock oscillates more rapidly.}
\label{bISPEC4}
\end{figure}

The change in behavior around $\mu\sim2|\lambda|$, for example, near $\mu_c\sim2.5$ when $|\lambda|\sim1$, should be interpreted as a crossover in the Whittaker connection coefficients rather than as a new feature in the momentum dependence of the bispectrum. For fixed $\mu$ and $\lambda$, the squeezed-limit signal remains a logarithmic clock signal. The tilted deformation modifies the magnitude $|C_+|$ and the phase $\arg \left(C_+\right)$, but it does not generate any kink or transition as a function of $\kappa=k_S/k_L$. Nevertheless, the phase of the oscillatory signal may itself encode additional physical information beyond the clock frequency~\cite{Qin:2022xjq}. For fixed $\mu$, the tilt leaves the oscillation period unchanged while shifting $\arg \left(C_+\right)$, see Fig.~\ref{bISPEC4}. Equivalently, the tilted and pure-ghost signals are horizontally displaced relative to one another in $\log\kappa$-space. Depending on the sign and magnitude of $\lambda$, this displacement can make the tilted contribution appear approximately in phase or out of phase with the pure-ghost signal.

\subsection{Comparison with Whittaker modes in the literature}
\label{subsec:chemical_comparison}

\begin{figure}[ht]
\hfill
{\includegraphics[width=7cm]{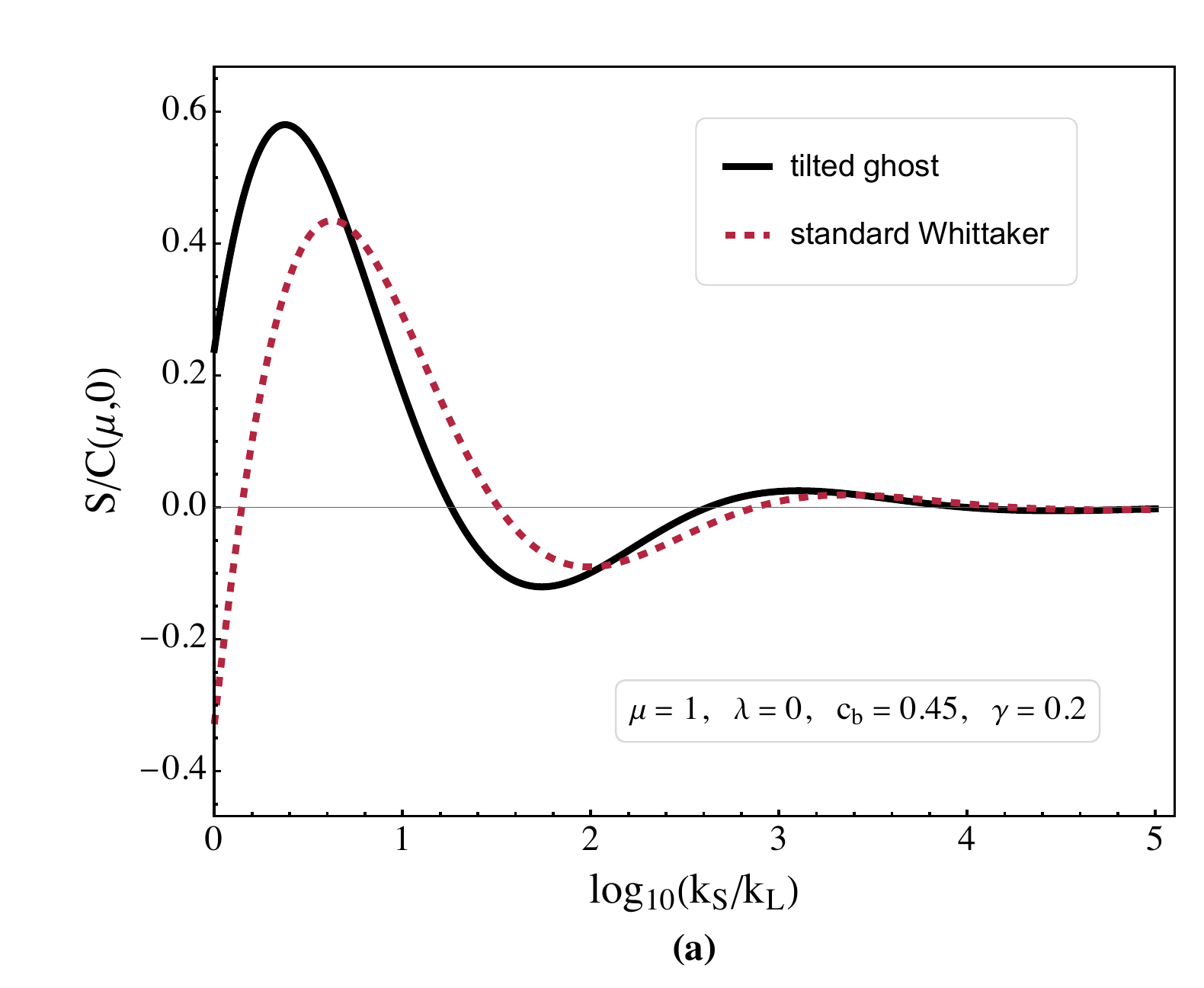}}
\hfill
{\includegraphics[width=7cm]{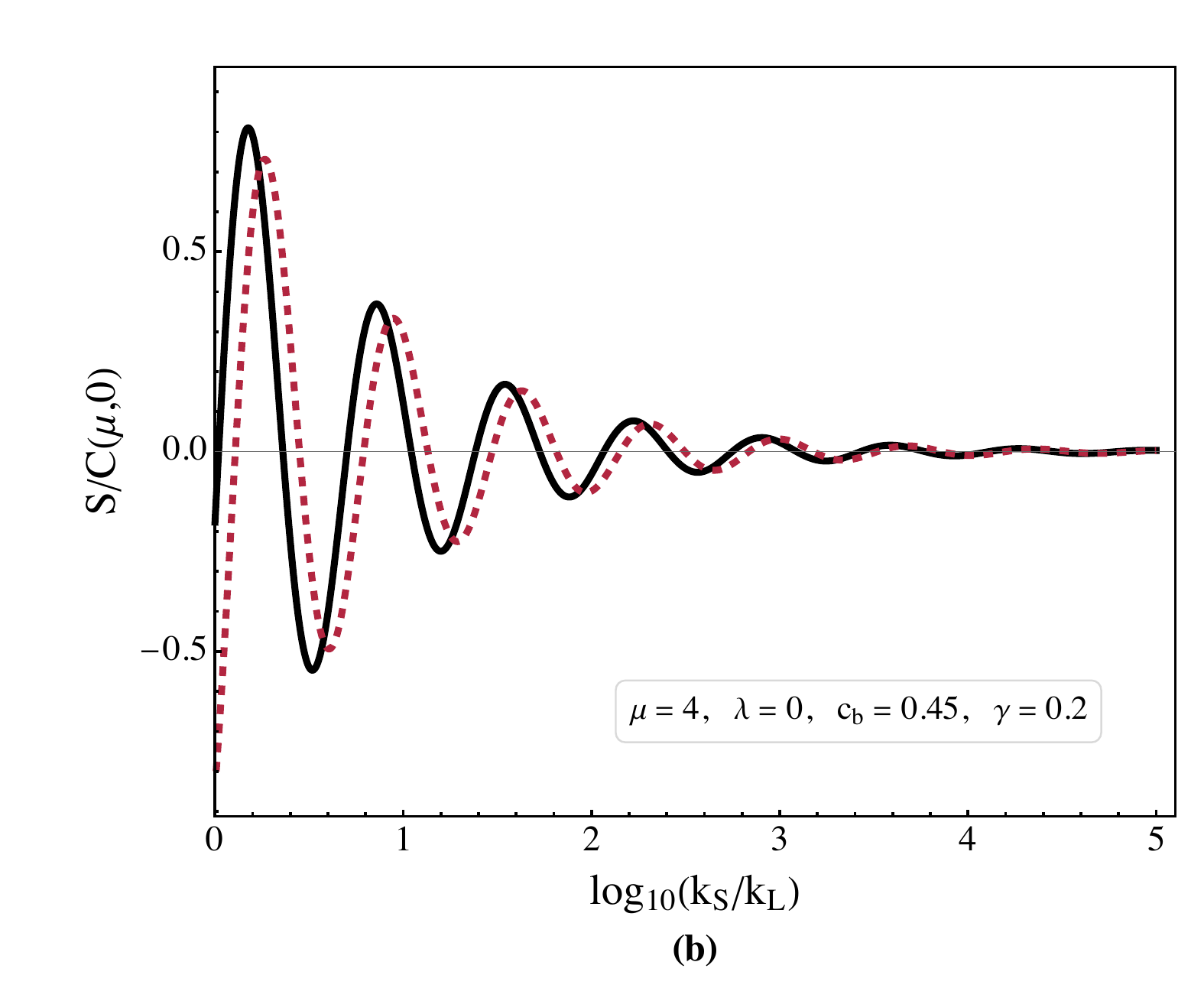}}
\hfill
\caption{Comparison between the normalized tilted-ghost signal and the standard Whittaker signal in the undeformed case \(\lambda=0\), shown as a function of \(\log_{10}(k_S/k_L)\). The solid black curve corresponds to the tilted-ghost spectator, while the dashed red curve corresponds to the standard Whittaker result with effective sound speed \(c_b\simeq\sqrt\gamma\). In both panels the signal is normalized by the corresponding \(\lambda=0\) amplitude, \(S/C(\mu,0)\), and we fix \(\gamma=0.20\). Panel (a) shows \(\mu=1\), while panel (b) shows \(\mu=4\). The remaining difference at \(\lambda=0\) reflects the fact that the two templates reduce to different ultraviolet vacua: pure-ghost WKB for the tilted-ghost mode and relativistic Bunch--Davies/Hankel for the standard Whittaker mode.}
\label{fig:lambda-zero-mu1-mu4-comparison}
\end{figure}

We now compare the tilted-ghost result with the standard Whittaker benchmark commonly used in the boostless and chemical-potential cosmological-collider literature. For the comparison below, we set $c_b^2\simeq\gamma$ only as a matching prescription between the dimensionless clock arguments of the two templates. This should not be interpreted as a physical identification between the standard Whittaker sound speed $c_b$ and the tilted-ghost parameter $\gamma=H/M$. The former characterizes the low-speed propagation of the heavy sector in the standard boostless template, whereas the latter measures the ratio between the Hubble scale and the ghost scale controlling the quartic dispersion. With this choice, the leading logarithmic arguments are aligned, and the remaining differences more directly reflect the distinct Whittaker coefficients and ultraviolet normalizations.
Figure~\ref{fig:lambda-zero-mu1-mu4-comparison} shows that the two templates do not coincide exactly, even if $\lambda=0$. This is expected since the $\lambda=0$ limit of the tilted-ghost solution reduces to the pure-ghost mode, whereas the $\lambda=0$ limit of the conventional Whittaker template corresponds to the relativistic Hankel/Bunch-Davies mode. The residual difference at $\lambda=0$ therefore already reflects a distinction in the ultraviolet propagation regime and, consequently, in the underlying vacuum prescription.

\begin{figure}[ht]
\centering
\begin{minipage}{0.48\linewidth}
\centering
{\includegraphics[width=\linewidth]{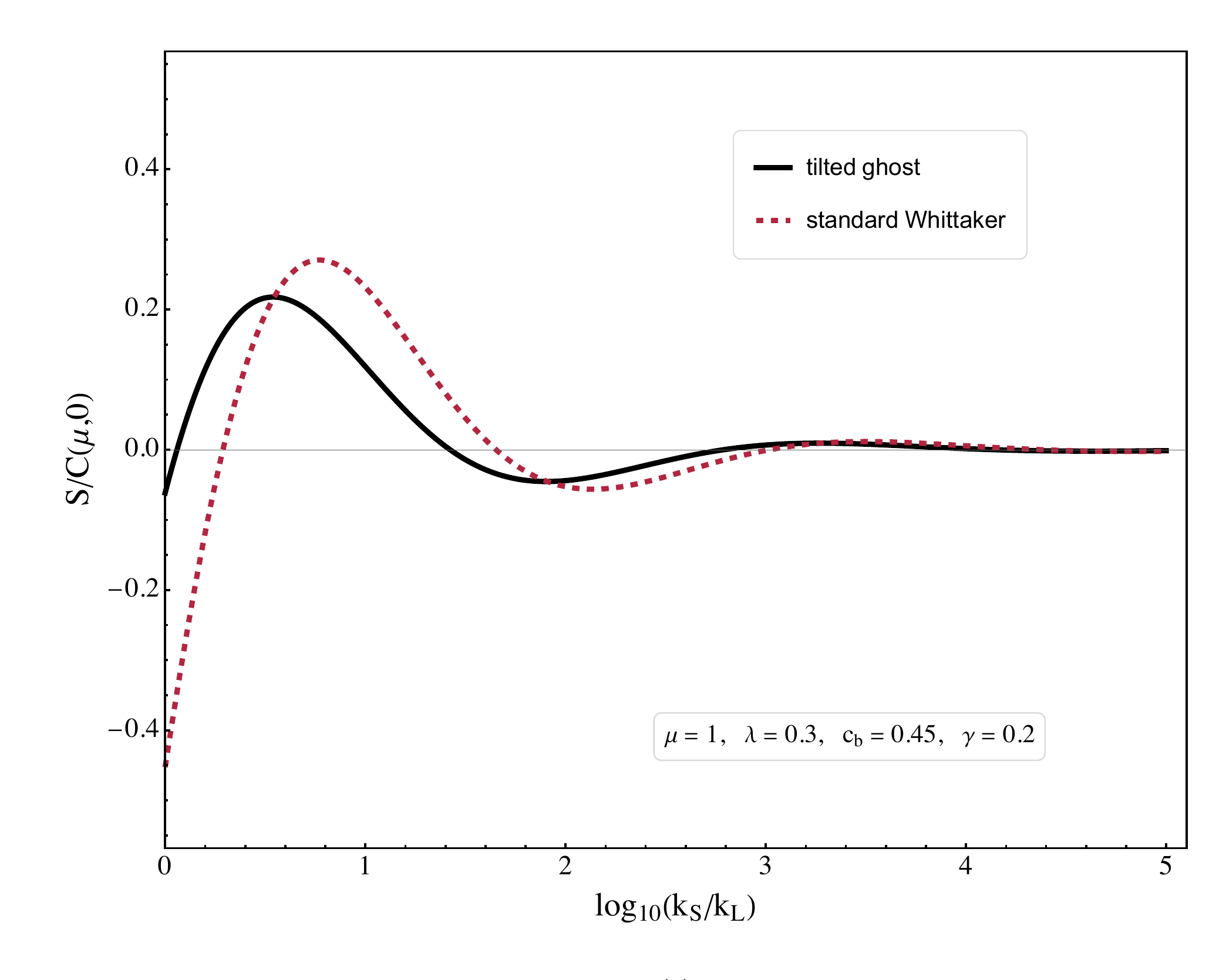}}
\end{minipage}\hfill
\begin{minipage}{0.48\linewidth}
\centering
{\includegraphics[width=\linewidth]{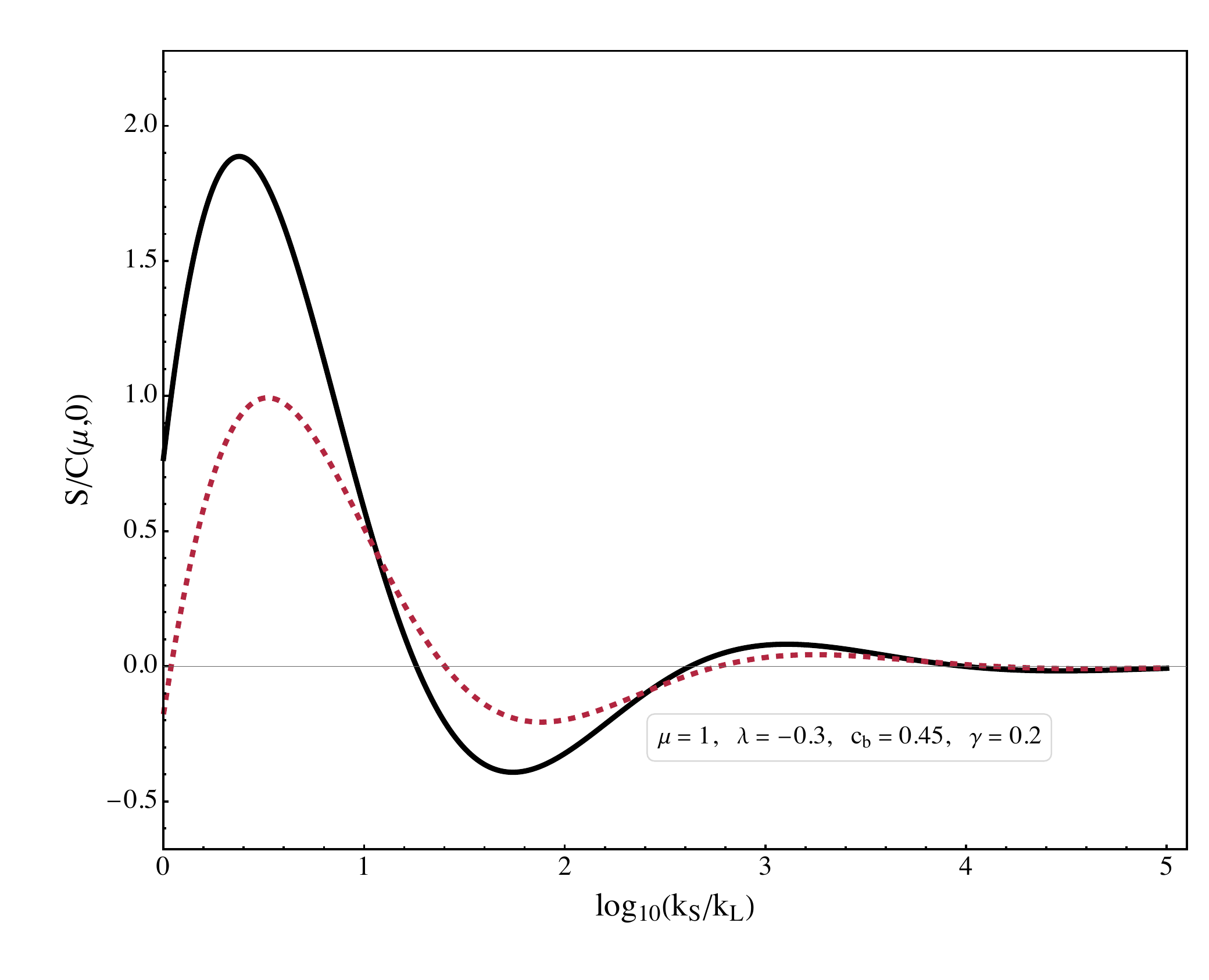}}
\end{minipage}
\vspace{0.3cm}
\begin{minipage}{0.48\linewidth}
\centering
{\includegraphics[width=\linewidth]{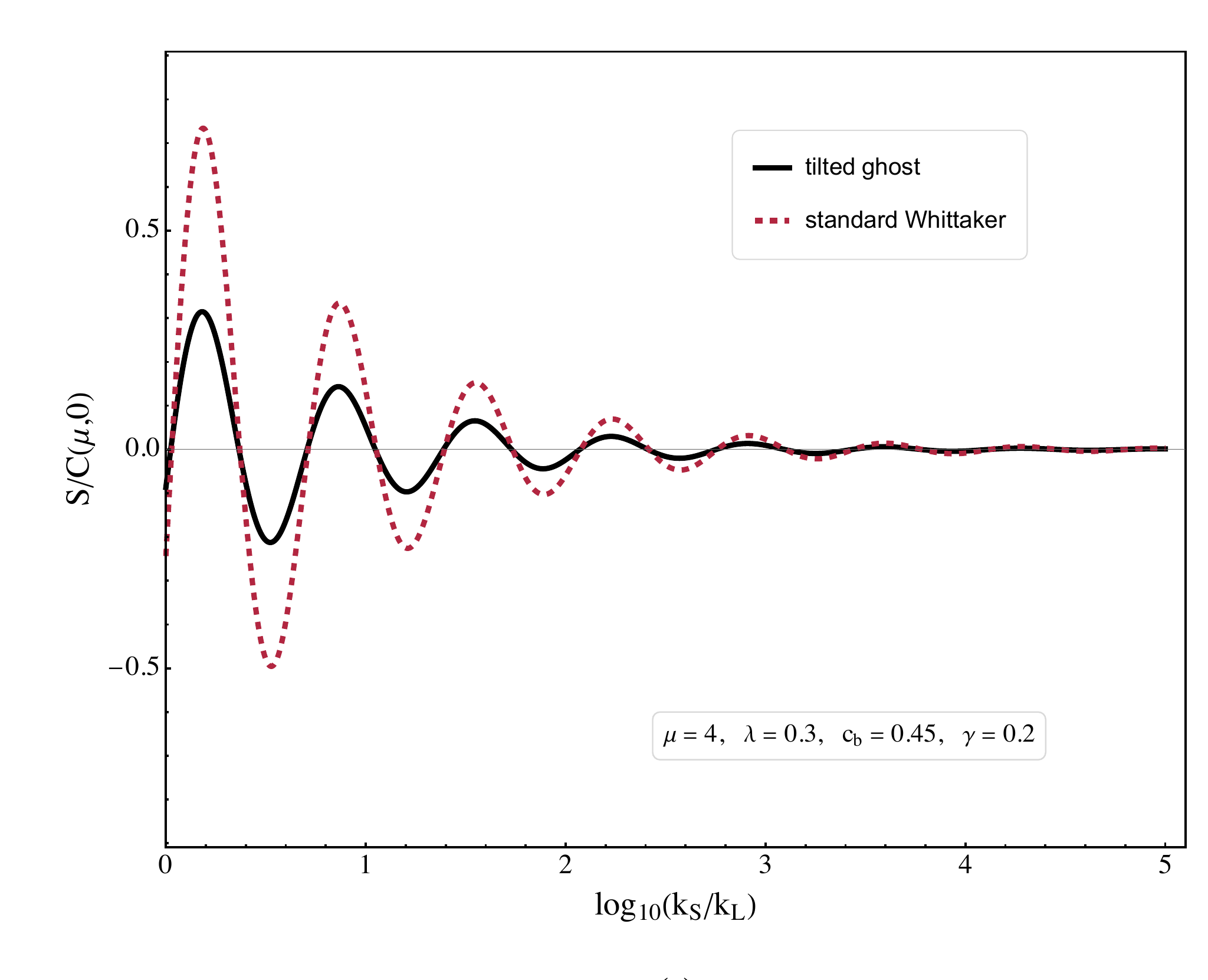}}
\end{minipage}\hfill
\begin{minipage}{0.48\linewidth}
\centering
{\includegraphics[width=\linewidth]{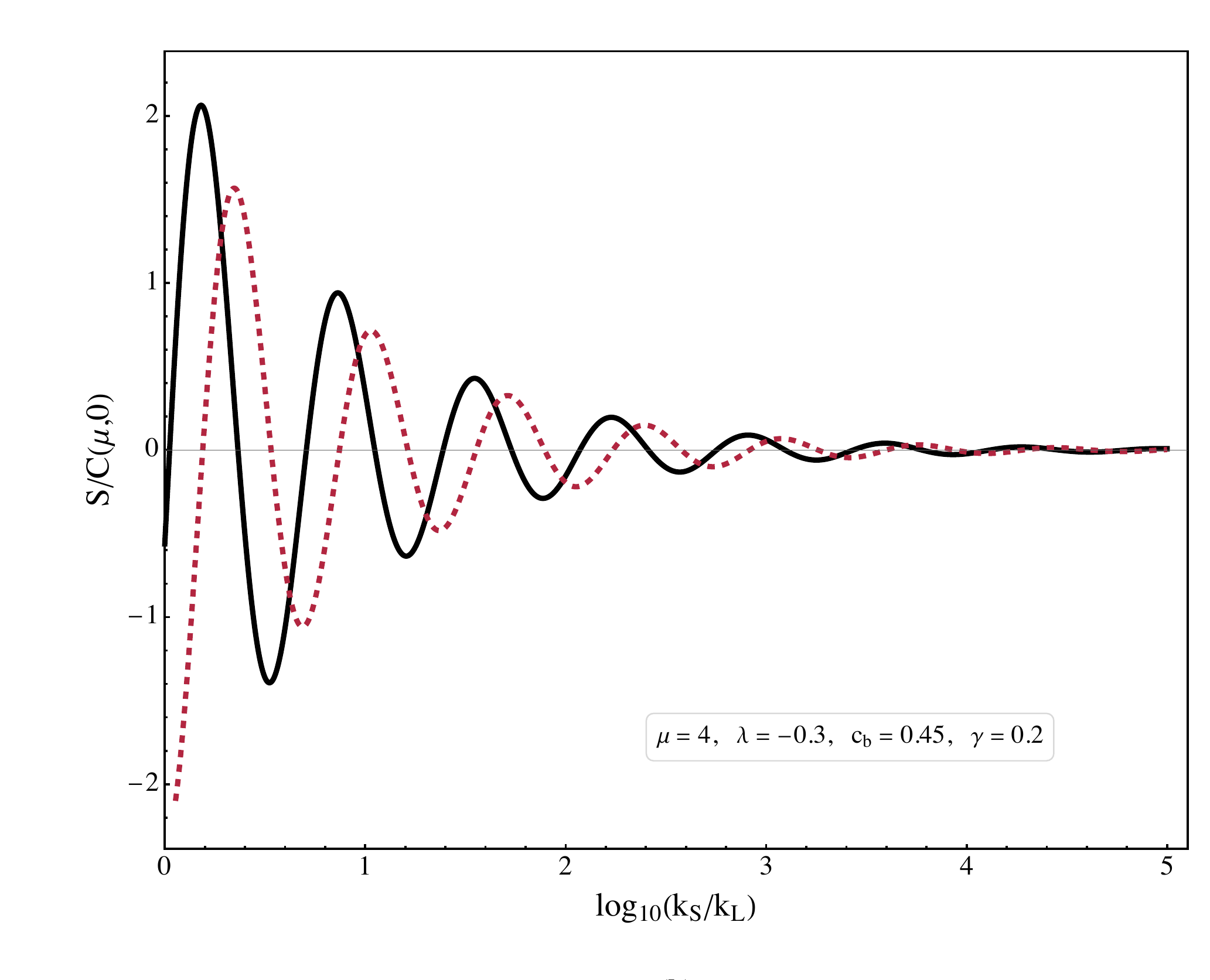}}
\end{minipage}
\caption{Comparison between the normalized tilted-ghost signal and the standard Whittaker signal in the squeezed configuration. The solid black curves correspond to the tilted-ghost spectator, while the dashed red curves show the standard Whittaker result with effective sound speed \(c_b\simeq\sqrt\gamma\). The vertical axis is normalized by the corresponding \(\lambda=0\) amplitude, \(S/C(\mu,0)\), which removes the large overall normalization difference associated with the different initial conditions. The upper row shows \(\mu=1\) and the lower row shows \(\mu=4\); the left column has \(\lambda=0.3\) and the right column has \(\lambda=-0.3\). The remaining differences therefore isolate the propagation-induced Whittaker coefficients rather than a trivial mismatch of vacuum normalization.}
\label{fig:tilted-ghost-vs-standard-whittaker-lambda-sign}
\end{figure}

For nonzero $\lambda$, Fig.~\ref{fig:tilted-ghost-vs-standard-whittaker-lambda-sign} shows that the two Whittaker realizations respond differently to the sign of the deformation. In the tilted-ghost case, the signs of $\lambda$ entering the late-time coefficients are reversed relative to those appearing in the standard Whittaker coefficients. As a consequence, the $\lambda>0$ branch becomes more strongly Boltzmann suppressed, whereas the $\lambda<0$ branch is enhanced relative to the corresponding standard Whittaker benchmark. The asymmetric dependence on the sign of $\lambda$ constitutes the distinctive phenomenological imprint of the tilted-ghost origin of the Whittaker equation.

The comparison reveals an important phenomenological aspect of the tilted-ghost scenario. The large amplitude differences observed prior to normalization are not determined exclusively by the late-time clock dynamics but depend sensitively on the underlying choice of initial conditions. The tilted-ghost mode is Bunch-Davies-like only in an adiabatic WKB sense: it selects a positive-frequency branch in the asymptotic past, yet the corresponding frequency is governed by the non-relativistic $k^4$ dispersion relation rather than by the relativistic $k^2$ behavior characteristic of the standard de Sitter vacuum. 

Consequently, it does not coincide with the conventional de Sitter Bunch-Davies prescription. This distinction is especially important when comparing the tilted-ghost signal with recent mechanisms designed to extend the mass reach of the cosmological collider~\cite{Bodas_2021,Bodas:2024hih,Bodas:2025chargedloops,Aoki:2026modularCP}. In ordinary de Sitter exchange, the non-analytic signal of a principal-series particle is accompanied by the standard Boltzmann factor $e^{- \pi \mu}$. Thus, for \(m/H\gg1\), the clock signal is exponentially suppressed, even though the logarithmic oscillation remains present. The recent chemical-potential constructions evade this conclusion by introducing an additional energy scale in the heavy sector. For scalar and vector chemical potential models, the rolling inflaton background induces a coupling of the schematic form \(\dot\phi_0 J^0/\Lambda\), so that the chemical potential can source heavy particles efficiently whenever their mass lies below the chemical potential scale \cite{Bodas_2021}. In this case, the relevant condition for avoiding the usual Boltzmann suppression is approximately
\begin{equation}
\mu \lesssim \lambda_{\rm ch},
\qquad
\lambda_{\rm ch}\lesssim \dot\phi_0^{1/2}\sim 60H .
\end{equation}
For charged particles appearing only in loops~\cite{Bodas:2025chargedloops}, the analogous condition is imposed on the pair-production threshold, schematically \(\mu_1+\mu_2\lesssim \widehat\lambda_{\rm ch}\).~Moreover, the late-time mode functions in Bodas~et~al. are derived within a WKB treatment and do not retain the standard $\Gamma$-function structure that underlies the analytic amplitude estimates commonly used in conventional cosmological-collider analyses. The enhancement is not only determined by a shift of the oscillatory phase: the magnitude of the relevant connection coefficients can also modify the signal, allowing a compensation between the Boltzmann factor and substantially extend the accessible mass range \cite{Ferreira:2026ghostspectators,Yin:2023jlv}.

Non-Bunch–Davies initial states provide a different route~\cite{Yin:2023jlv}: the suppression can be canceled by the initial state structure of the propagator rather than by energy injection from a chemical potential. It is instructive to compare this mechanism with the non-Bunch--Davies $\alpha$-vacuum scenario of \cite{Yin:2023jlv}. In that case, the leading non-BD seed contributions cancel the usual Hubble-scale Boltzmann factor, so that their amplitude is controlled by the Bogoliubov mixing, schematically $\sinh\alpha$, rather than by $e^{-\pi\mu}$. By contrast, a positive tilted-ghost deformation, enhances the pure-ghost signal but does not generically eliminate its residual mass dependence.

At the level of the oscillatory envelope, a representative comparison between the $\alpha$-vacuum benchmark $\alpha=1$, $z_\Lambda/H=20$, and a positive-tilt benchmark with $\gamma=10^{-2}$, gives
\begin{align}
&\mu = 1: \quad \alpha =1,\quad f_{\rm NL}^{\rm osc} \sim {\cal O}(10^{-1}); \qquad  \lambda = -1, \quad f_{\rm NL}^{\rm osc} \sim {\cal O}(1).\\ 
&\mu = 8:\quad \,  \alpha =1, \quad f_{\rm NL}^{\rm osc} \sim {\cal O}(10^{-2}); \qquad \lambda = -1, \quad f_{\rm NL}^{\rm osc} \sim {\cal O}(10^{-5}).
\end{align}
Thus, for moderately heavy fields, the positive-tilt signal can be comparable to, or somewhat larger than, its $\alpha$-vacuum counterpart. For substantially heavier fields, however, the non-Bunch--Davies mechanism remains parametrically more efficient: even with positive tilt, the tilted-ghost amplitude is separated by at least three orders of magnitude in this representative comparison. The precise numerical ratio is not universal, since it depends on the interaction normalization and on the cutoff prescription in the $\alpha$-vacuum setup.

This hierarchy also distinguishes the tilted-ghost scenario from chemical-potential constructions. In the latter, an independent background-induced scale can compensate for the Boltzmann factor parametrically when it exceeds the particle mass. The $\alpha$-vacuum instead removes the suppression through initial-state Bogoliubov mixing, whereas tilted ghosts modify the late-time branch weights through non-relativistic propagation and the associated Whittaker connection coefficients. Consequently, the tilted-ghost mechanism can substantially improve the visibility of heavy particles relative to ordinary de Sitter exchange, but it does not generically provide the unrestricted high-mass reach of either non-Bunch--Davies or chemical-potential scenarios.

The tilted-ghost mechanism is qualitatively different from all these cases. There is no conserved charge, no helicity asymmetry, and no external chemical-potential that overcomes the mass scale. The exchanged field is a scalar spectator whose propagation is governed by a higher-derivative dispersion relation. The suppression is weakened because the mode does not propagate as a relativistic de Sitter mode in the ultraviolet. In particular, the initial condition imposed in the ultraviolet changes the amplitude of the signal. For example, one finds
\begin{equation}
\frac{{\cal A}_{\rm ghost}(\mu,\gamma)}
     {{\cal A}_{\rm dS}(\mu)}
=\left.
\frac{\left|\mathcal C_+^{\rm G}\right|}
     {\left|\mathcal C_+^{\rm dS}\right|}
\right|_{\gamma=10^{-3},\,\mu=8}
\sim
{\cal O}(10^6).
\end{equation}
Therefore, for $\lambda\sim \mp \, O(1)$, this amplitude can be further enhanced or suppressed by one or two orders of magnitude, depending on the sign of the tilt. It is also useful to compare this with the tilted-ghost amplitude relative to the standard de Sitter result,
\begin{equation}
\frac{{\cal A}_{\rm TG}(\mu,\lambda,\gamma)}
     {{\cal A}_{\rm dS}(\mu)}
=\left.
\frac{\left|\mathcal C_+^{\rm TG}\right|}
     {\left|\mathcal C_+^{\rm dS}\right|}
\right|_{\mu=1,\,\lambda=-1,\,\gamma=10^{-2}}
\sim
{\cal O}(10^2).
\end{equation}
The oscillatory signal from the exchange of tilted modes with relatively small masses, for instance $\mu=1$ or $\mu=2$, can reach the interval
\begin{equation}
f_{\rm NL}^{\rm osc}\sim {\cal O}(0.01-1).
\label{eq:fNLestimate}
\end{equation}
This estimate is meant as a benchmark normalization for the non-analytic clock signal. In the conventions used here, the dependence on the interaction strengths enters through the dimensionless combination
\begin{equation}
f_{NL}^{osc}
\sim  
\left(\frac{\rho}{H}\right)(gH) \, \mathcal{F}_{TG}(\mu,\gamma,\lambda)
\end{equation}
multiplying the Whittaker coefficient that controls the soft non-analytic contribution. The range in Eq.~\eqref{eq:fNLestimate} corresponds to perturbative values of the interaction factor, typically ${\cal A}_{\rm int}\sim 10^{-2}-10^{-1}$, together with light principal-series masses, $\mu={\cal O}(1)$, and a controlled signed tilt, $|\lambda|\lesssim 1$. Values near the upper end of the interval are obtained on the enhanced branch, $\lambda<0$, or for moderately larger but still perturbative interaction factors. Other choices of $\rho$ and $g$ rescale the estimate proportionally.

However, the window $| \lambda |>\mu$ is rather restrictive unless the asymmetry scale is large, which limits the range of masses that can effectively escape the Boltzmann suppression.

The degree of non-linearity can affect the amount of enhancement of the signal. Beyond the pure-ghost and tilted-ghost spectators considered here, one may contemplate more general modes; for example, those arising in Ho\v{r}ava--Lifshitz-inspired settings~\cite{Zhu_2014,Mukohyama:2009zs}, with dispersion relations of the schematic form:
\begin{equation}
\omega_k(\eta)^2 \propto k^2+\eta^2 k^4+\eta^4 k^6 
\quad .
\end{equation}
Such modes may provide a different mass range in which the usual Boltzmann suppression is softened. In this sense, 
the tilted-ghost case may be viewed as a special case of a broader class of non-linear collider signals; we will investigate them in future work. The pure-ghost dispersion already changes the usual de Sitter suppression from \(e^{-\pi\mu}\) to a softer behavior of the form \(e^{-\pi\mu/2}\). The tilted deformation then modifies this result through the sign-dependent Whittaker coefficients. It is therefore natural to ask whether higher powers in the dispersion relation can soften the suppression even further.
This also elucidates the status of the tilted-ghost model relative to chemical-potential and non-Bunch--Davies scenarios. As in those cases, the tilted-ghost signal can make heavier particles more visible than in ordinary de Sitter exchange. However, the mechanism is different. There is no external scale that overcomes the mass scale, and therefore the model does not generically imply a parametric reach up to \(m\sim 60H\). The enhancement is instead controlled by the non-relativistic propagation of the spectator and by the late-time Whittaker connection coefficients.

The same distinction can be stated in terms of branch weights. In chemical-potential cosmological-collider models, the Whittaker parameter is generated by the coupling between the rolling inflaton background and a current in the heavy sector, which may distinguish particles from antiparticles or different helicity states. In the tilted-ghost case, there is no conserved charge, no helicity structure, and no external chemical-potential scale. The exchanged field is a scalar spectator, and the parameter \(\lambda\) only controls how the pure-ghost WKB mode is projected onto the two late-time branches. Thus, \(\lambda\) acts as a chemical-potential-like parameter, but its origin is the propagation of the mode itself. This is also why absolute comparisons with the standard Whittaker template are sensitive to the ultraviolet normalization: the tilted-ghost mode is normalized in the pure-ghost regime, whereas the standard Whittaker mode is normalized in a relativistic Bunch--Davies regime. After normalizing each signal by its corresponding \(\lambda=0\) amplitude, this large vacuum-normalization effect is removed, and the remaining physical information is the sign-asymmetric change in the envelope and the phase shift of the clock signal. Therefore, the tilted-ghost Whittaker mode should not be interpreted as the usual chemical-potential Whittaker mode written with different parameters. Both cases are described by Whittaker functions, and both modify the non-analytic oscillatory signal, but the tilted-ghost signal has a different ultraviolet prescription, a different \(\lambda=0\) limit, and a different response under \(\lambda\to-\lambda\).

\section{Conclusions}
\label{sec:conclusions}

We have studied a non-relativistic realization of cosmological-collider physics in which the exchanged heavy field is a tilted-ghost spectator. The central distinction from more conventional boostless or chemical-potential templates is that the deformation appears already in the free propagation of the spectator. The ultraviolet branch is selected by the pure-ghost WKB regime, while the quadratic tilt deforms the exact Whittaker mode and changes the late-time connection coefficients.

The exact Whittaker solution makes the relevant physical regimes transparent. For light modes, the positive-tilt branch can pass through a transient unstable interval before the field approaches its frozen late-time behavior. For principal-series modes, the same crossover between the quartic and quadratic terms persists, but in the controlled regime $\beta<2\gamma\sqrt{\mu^2+1/4}$, it does not correspond to a real turning point. Instead, the tilt is encoded in the relative weights and phases of the two oscillatory branches $( -\eta)^{3/2\pm i\mu}$. This is the precise sense in which $\lambda=\lambda_{\rm eff}$ acts as a chemical-potential-like parameter. Using the Schwinger-Keldysh formalism, we isolated the soft exchange channel responsible for the non-analytic squeezed-limit signal. The resulting clock signal retains the standard logarithmic frequency $\mu$, but its amplitude and phase are dressed by the tilted-ghost Whittaker coefficients. The hard exchange channels remain analytic in the squeezed ratio and therefore contribute to the background rather than to the non-analytic clock.

The phenomenological comparison with the standard Whittaker modes in the literature shows that the tilted-ghost result is not obtained by a simple reparameterization of the usual chemical-potential template. The two signals have the same clock exponent but different Whittaker indices, different ultraviolet prescriptions, and different limiting modes: the standard Whittaker template reduces to the relativistic Hankel/Bunch–Davies mode, whereas the tilted-ghost template reduces to the pure-ghost WKB mode. This difference is visible already at $\lambda=0$, and it becomes sharper for nonzero tilt. In particular, the signs of the $\lambda$-dependent coefficients imply an asymmetric response: the $\lambda>0$ branch is comparatively Boltzmann suppressed, while the $\lambda<0$ branch is enhanced.

A central feature emerging from the normalized plots is that the largest amplitude differences are intimately connected to the choice of initial condition. Once each signal is normalized by its corresponding $\lambda=0$ amplitude, the large vacuum-normalization effect is removed, leaving as the most robust tilted-ghost signatures the sign-asymmetric modulation of the envelope and the phase displacement of the clock signal. This highlights that the vacuum prescription is itself an intrinsic part of the phenomenology: the tilted mode is Bunch-Davies-like only in the adiabatic WKB sense appropriate to a $k^4$ dispersion relation, rather than in the standard relativistic de Sitter sense.

Several extensions of the present analysis would be valuable. First, the allowed parameter range should be embedded within a complete effective field theory framework in which the small-tilt condition, as well as the possible emergence of additional turning-point windows, can be analyzed systematically. Second, the same propagation-induced mechanism should be investigated in higher-point observables, particularly in the collapsed trispectrum and in bootstrap formulations where the singularity structure associated with the exchanged mode becomes explicit. Finally, the phase and amplitude shifts identified here could be incorporated into observational templates in order to determine under which conditions tilted-ghost exchange can be distinguished from standard boostless, pure-ghost, or chemical-potential cosmological-collider signals, in line with the recent transition from analytic collider templates to concrete CMB and large-scale structure searches.

\acknowledgments 
The authors acknowledge the financial support of the National Scientific and Technological Research Council (CNPq, Brazil) and the Brazilian Federal Agency for Support and Evaluation of Graduate Education (CAPES, Brazil).


\clearpage

\appendix

\section{Schwinger--Keldysh Computations}
\label{app:sk_exchange}

In this appendix, we spell out the Schwinger–Keldysh structure of the exchange
contribution to the scalar three-point function. The goal is not to emphasize
the final closed-form result, but rather to make explicit how the different SK
components are organized, how the soft and hard channels are separated, and how
the non-analytic squeezed-limit contribution is extracted. We start from the exchange contribution
\begin{align}
\big\langle
\varphi_{\mathbf k_1}
\varphi_{\mathbf k_2}
\varphi_{\mathbf k_3}
\big\rangle'_{\sigma}
=
\mathcal C_{12|3}
+
\mathcal C_{23|1}
+
\mathcal C_{31|2}.
\label{app:eq:channel_decomposition}
\end{align}
The notation \(\mathcal C_{ij|\ell}\) means that the two hard external legs
\((i,j)\) meet at one vertex, while the leg \(\ell\) meets the exchanged field
at the second vertex. The exchanged momentum in this channel is \(s_{ij}\).

\subsection{Soft channel and Schwinger-Keldysh components}

In the squeezed limit $k_3\ll k_1\sim k_2$, the channel \(\mathcal C_{12|3}\) is singled out because $s_{12}=|\mathbf k_1+\mathbf k_2|=k_3\to 0$. This is the only exchange channel in which the small-momentum expansion of the internal propagator can be used. Defining
$K\equiv k_1+k_2$, we may write
\begin{align}
\mathcal C_{12|3}
=
\frac{H}{4k_1k_2k_3}
\sum_{\epsilon,\epsilon'=\pm1}\epsilon\epsilon'\,
\mathcal I^{\epsilon\epsilon'}_{12|3},
\end{align}
with
\begin{align}
\mathcal I^{\epsilon\epsilon'}_{12|3}
=
\int_{-\infty}^{0}d\eta_1
\int_{-\infty}^{0}\frac{d\eta_2}{(-\eta_2)^2}\,
e^{i\epsilon K\eta_1+i\epsilon' k_3\eta_2}
G_{\epsilon\epsilon'}(k_3,\eta_1,\eta_2).
\label{app:eq:Iab_soft_def}
\end{align}

The four Schwinger-Keldysh propagators are expressed in terms of the Wightman function
\(\mathcal G(k;\eta_1,\eta_2)\) as
\begin{align}
G_{-+}(k;\eta_1,\eta_2)
&=
\mathcal G(k;\eta_1,\eta_2),
\\
G_{+-}(k;\eta_1,\eta_2)
&=
\mathcal G(k;\eta_2,\eta_1),
\\
G_{++}(k;\eta_1,\eta_2)
&=
\Theta(\eta_1-\eta_2)\mathcal G(k;\eta_1,\eta_2)
+
\Theta(\eta_2-\eta_1)\mathcal G(k;\eta_2,\eta_1),
\\
G_{--}(k;\eta_1,\eta_2)
&=
\Theta(\eta_2-\eta_1)\mathcal G(k;\eta_1,\eta_2)
+
\Theta(\eta_1-\eta_2)\mathcal G(k;\eta_2,\eta_1).
\end{align}
Therefore, in the soft channel, we may replace the Wightman function with its
late-time expansion, $\mathcal G(k_3;\eta_1,\eta_2) \longrightarrow \mathcal G_{\rm LT}(k_3;\eta_1,\eta_2)$. The corresponding four contour integrals are
\begin{align}
\mathcal I^{+-}_{12|3}
&=
\int_{-\infty}^{0}d\eta_1
\int_{-\infty}^{0}\frac{d\eta_2}{(-\eta_2)^2}\,
e^{iK\eta_1-ik_3\eta_2}
\mathcal G_{\rm LT}(k_3;\eta_2,\eta_1),
\\
\mathcal I^{-+}_{12|3}
&=
\int_{-\infty}^{0}d\eta_1
\int_{-\infty}^{0}\frac{d\eta_2}{(-\eta_2)^2}\,
e^{-iK\eta_1+ik_3\eta_2}
\mathcal G_{\rm LT}(k_3;\eta_1,\eta_2),
\\
\mathcal I^{++}_{12|3}
&=
\int_{-\infty}^{0}d\eta_1
\int_{-\infty}^{0}\frac{d\eta_2}{(-\eta_2)^2}\,
e^{iK\eta_1+ik_3\eta_2}
\Big[
\Theta(\eta_1-\eta_2)\mathcal G_{\rm LT}(k_3;\eta_1,\eta_2)
\nonumber\\
&\hspace{5.1cm}
+
\Theta(\eta_2-\eta_1)\mathcal G_{\rm LT}(k_3;\eta_2,\eta_1)
\Big],
\\
\mathcal I^{--}_{12|3}
&=
\int_{-\infty}^{0}d\eta_1
\int_{-\infty}^{0}\frac{d\eta_2}{(-\eta_2)^2}\,
e^{-iK\eta_1-ik_3\eta_2}
\Big[
\Theta(\eta_2-\eta_1)\mathcal G_{\rm LT}(k_3;\eta_1,\eta_2)
\nonumber\\
&\hspace{5.1cm}
+
\Theta(\eta_1-\eta_2)\mathcal G_{\rm LT}(k_3;\eta_2,\eta_1)
\Big].
\label{app:eq:soft_SK_components}
\end{align}

The important bookkeeping point is that the mixed late-time branches are not
symmetric under \(\eta_1\leftrightarrow\eta_2\). Hence, one must keep track of
whether the propagator appears as \(\mathcal G_{\rm LT}(k_3;\eta_1,\eta_2)\) or as
\(\mathcal G_{\rm LT}(k_3;\eta_2,\eta_1)\). We use $\Delta_\pm \equiv 3/2\pm i\mu $. The late-time propagator can be decomposed as
\begin{align}
\mathcal G_{\rm LT}(k_3;\eta_1,\eta_2)
&= \mathcal A_-\,k_3^{-2i\mu} (-\eta_1)^{\Delta_-} (-\eta_2)^{\Delta_-} +\mathcal B_{-+} (-\eta_1)^{\Delta_-} (-\eta_2)^{\Delta_+}
\nonumber\\
&\quad + \mathcal A_+\,k_3^{2i\mu} (-\eta_1)^{\Delta_+} (-\eta_2)^{\Delta_+} + \mathcal B_{+-} (-\eta_1)^{\Delta_+} (-\eta_2)^{\Delta_-}\qquad .
\label{app:eq:GLT_decomp}
\end{align}
The coefficients are
\begin{align}
\mathcal A_- 
&= \frac{H^2}{2e^{\pi\lambda}} \gamma^{-i\mu} \frac{\Gamma(i\mu)^2} { \Gamma\!\left(\frac12-i\lambda+\frac{i\mu}{2}\right) \Gamma\!\left(\frac12+i\lambda+\frac{i\mu}{2}\right) }
\quad , \\
\mathcal A_+
&= \frac{H^2}{2e^{\pi\lambda}} \gamma^{i\mu} \frac{\Gamma(-i\mu)^2} { \Gamma\!\left(\frac12-i\lambda-\frac{i\mu}{2}\right) \Gamma\!\left(\frac12+i\lambda-\frac{i\mu} {2}\right) }
\quad ,\\
\mathcal B_{-+}
&=\frac{H^2}{2\mu\,e^{\pi(\lambda+\mu/2)}}\operatorname{csch}(\pi\mu)\cosh\!\left(\pi\lambda-\frac{\pi\mu}{2}\right)
\quad ,\\
\mathcal B_{+-}
&=\frac{H^2}{2\mu\,e^{\pi(\lambda-\mu/2)}}\operatorname{csch}(\pi\mu)\cosh\!\left(\pi\lambda+\frac{\pi\mu}{2}\right)
\quad .
\end{align}
For real \(\lambda\) and \(\mu\), $\mathcal A_+=\mathcal A_-^*$, $\mathcal B_{-+},\mathcal B_{+-}\in\mathbb R$, and $\mathcal B_{-+}\neq \mathcal B_{+-}$. To evaluate the time integrals, it is useful to first consider a generic
monomial
\begin{align}
\mathcal G_{p,q}(\eta_1,\eta_2)
=
\mathcal C\,(-\eta_1)^p(-\eta_2)^q .
\label{app:eq:generic_monomial}
\end{align}
The four terms in \eqref{app:eq:GLT_decomp} are then obtained by the
substitutions
\begin{align}
(p,q)
=
(\Delta_-,\Delta_-),
\qquad
(\Delta_-,\Delta_+),
\qquad
(\Delta_+,\Delta_+),
\qquad
(\Delta_+,\Delta_-),
\end{align}
together with the appropriate coefficients and explicit powers of \(k_3\). The Gamma functions arise from the elementary contour integrals
\begin{align}
J_\nu^{(-)}(A)
&\equiv
\int_{-\infty(1-i\epsilon)}^{0}
d\eta\,
e^{+iA\eta}
(-\eta)^{\nu-1},
\\
J_\nu^{(+)}(A)
&\equiv
\int_{-\infty(1+i\epsilon)}^{0}
d\eta\,
e^{-iA\eta}
(-\eta)^{\nu-1},
\qquad A>0.
\end{align}
Setting \(\eta=-x\), with \(x>0\), gives
\begin{align}
J_\nu^{(-)}(A)
&=
\int_0^\infty dx\,
x^{\nu-1}
e^{-iAx}
e^{-\epsilon A x}
=
e^{-i\pi\nu/2}
\Gamma(\nu)
A^{-\nu},
\\
J_\nu^{(+)}(A)
&=
\int_0^\infty dx\,
x^{\nu-1}
e^{+iAx}
e^{-\epsilon A x}
=
e^{+i\pi\nu/2}
\Gamma(\nu)
A^{-\nu}.
\label{app:eq:basic_mellin_integrals}
\end{align}
These two identities are enough to evaluate the factorized Schwinger-Keldysh components and to extract the leading non-analytic part of the ordered components. For the \(+-\) component, using the monomial \eqref{app:eq:generic_monomial}, we find
\begin{align}
\mathcal I^{+-}[p,q]
&=
\mathcal C
\int_{-\infty}^{0}d\eta_1\,
e^{iK\eta_1}
(-\eta_1)^p
\int_{-\infty}^{0}
\frac{d\eta_2}{(-\eta_2)^2}\,
e^{-ik_3\eta_2}
(-\eta_2)^q .
\end{align}
With \(\eta_i=-x_i\), this becomes
\begin{align}
\mathcal I^{+-}[p,q]
&=
\mathcal C
\left[
\int_0^\infty dx_1\,
x_1^p e^{-iKx_1}
\right]
\left[
\int_0^\infty dx_2\,
x_2^{q-2} e^{+ik_3x_2}
\right].
\end{align}

Therefore,
\begin{align}
\mathcal I^{+-}[p,q]
&=-e^{i\pi(q-p)/2}\mathcal C\,\Gamma(p+1)\Gamma(q-1)\,K^{-1-p}k_3^{1-q}
\quad , \label{app:eq:Ipm_generic}\\
\mathcal I^{-+}[p,q]
&=-e^{-i\pi(q-p)/2}\mathcal C\,\Gamma(p+1)\Gamma(q-1)\,K^{-1-p}k_3^{1-q}
\quad .\label{app:eq:Imp_generic}
\end{align}
For any monomial
\((-\eta_1)^p(-\eta_2)^q\), the factorized Schwinger-Keldysh components generate the scaling $ K^{-1-p}k_3^{1-q}$. The \(++\) component contains time ordering:
\begin{align}
\mathcal I^{++}[p,q]
&=
\mathcal C
\int_{-\infty}^{0}d\eta_1
\int_{-\infty}^{0}
\frac{d\eta_2}{(-\eta_2)^2}\,
e^{iK\eta_1+ik_3\eta_2}
\nonumber\\&\quad\times
\Big[
\Theta(\eta_1-\eta_2)(-\eta_1)^p(-\eta_2)^q
+
\Theta(\eta_2-\eta_1)(-\eta_2)^p(-\eta_1)^q
\Big].
\end{align}
After setting \(\eta_i=-x_i\), one has $\Theta(\eta_1-\eta_2)=\Theta(x_2-x_1)$, $\Theta(\eta_2-\eta_1)=\Theta(x_1-x_2)$, and as a result we have
\begin{align}
\mathcal I^{++}[p,q]
&=
\mathcal C
\int_{0<x_1<x_2<\infty}dx_1dx_2
\Big[
e^{-iKx_1-ik_3x_2}
x_1^p x_2^{q-2}
+
e^{-iKx_2-ik_3x_1}
x_1^{p-2}x_2^q
\Big].
\label{app:eq:Ipp_ordered}
\end{align}
The second term is analytic in the squeezed ratio \(k_3/K\). The leading non-analytic contribution comes from the first term. Setting $x_1=u x_2$, $0<u<1$, one obtains
\begin{align}
\mathcal I^{++}_{\rm non-an.}[p,q]
&=
\mathcal C
\int_0^\infty dx_2\,
x_2^{p+q-1}
e^{-ik_3x_2}
\int_0^1du\,
u^p e^{-iKux_2}.
\end{align}
Performing the \(x_2\) integral first gives
\begin{align}
\mathcal I^{++}_{\rm non-an.}[p,q]
&=
\mathcal C
e^{-i\pi(p+q)/2}
\Gamma(p+q)
K^{-p-q}
\int_0^1du\,
u^p
\left(u+\frac{k_3}{K}\right)^{-p-q}.
\end{align}
The non-analytic squeezed contribution comes from the endpoint region
\(u\sim k_3/K\). Setting \(u=(k_3/K)t\), we find
\begin{align}
\int_0^1du\,
u^p
\left(u+\frac{k_3}{K}\right)^{-p-q}
&\simeq
\left(\frac{k_3}{K}\right)^{1-q}
\int_0^\infty dt\,
t^p(1+t)^{-p-q}
=
\left(\frac{k_3}{K}\right)^{1-q}
B(p+1,q-1).
\end{align}
Therefore
\begin{align}
\mathcal I^{++}_{\rm non-an.}[p,q]
&\simeq
e^{-i\pi(p+q)/2}
\mathcal C\,
\Gamma(p+1)\Gamma(q-1)
K^{-1-p}k_3^{1-q}.
\label{app:eq:Ipp_generic}
\end{align}
The same argument gives
\begin{align}
\mathcal I^{--}_{\rm non-an.}[p,q]
&\simeq
e^{+i\pi(p+q)/2}
\mathcal C\,
\Gamma(p+1)\Gamma(q-1)
K^{-1-p}k_3^{1-q}.
\label{app:eq:Imm_generic}
\end{align}
Equations \eqref{app:eq:Ipm_generic}, \eqref{app:eq:Imp_generic}, \eqref{app:eq:Ipp_generic}, and \eqref{app:eq:Imm_generic} are the basic Schwinger-Keldysh rules used to assemble the soft non-analytic contribution. The four monomials in the late-time propagator give two conjugate oscillatory
branches. Using
\begin{align}
\Delta_-=\frac32-i\mu,
\qquad
\Delta_+=\frac32+i\mu,
\end{align}
we repeatedly use
\begin{align}
\Gamma(\Delta_-+1)
=
\Gamma\!\left(\frac52-i\mu\right),
\qquad
\Gamma(\Delta_--1)
=
\Gamma\!\left(\frac12-i\mu\right),
\\
\Gamma(\Delta_++1)
=
\Gamma\!\left(\frac52+i\mu\right),
\qquad
\Gamma(\Delta_+-1)
=
\Gamma\!\left(\frac12+i\mu\right).
\end{align}

The scaling generated by the different monomials is as follows:
\begin{align}
\mathcal A_-\,k_3^{-2i\mu}
(-\eta_1)^{\Delta_-}
(-\eta_2)^{\Delta_-}
&\longrightarrow
K^{-5/2+i\mu}k_3^{-1/2-i\mu},
\\
\mathcal A_+\,k_3^{2i\mu}
(-\eta_1)^{\Delta_+}
(-\eta_2)^{\Delta_+}
&\longrightarrow
K^{-5/2-i\mu}k_3^{-1/2+i\mu},
\\
\mathcal B_{-+}
(-\eta_1)^{\Delta_-}
(-\eta_2)^{\Delta_+}
&\longrightarrow
K^{-5/2+i\mu}k_3^{-1/2-i\mu},
\\
\mathcal B_{+-}
(-\eta_1)^{\Delta_+}
(-\eta_2)^{\Delta_-}
&\longrightarrow
K^{-5/2-i\mu}k_3^{-1/2+i\mu}.
\end{align}
Thus the late-time propagator produces the two non-analytic squeezed branches
\begin{align}
K^{-5/2+i\mu}k_3^{-1/2-i\mu},
\qquad
K^{-5/2-i\mu}k_3^{-1/2+i\mu}.
\end{align}
Equivalently,
\begin{align}
K^{-5/2+i\mu}k_3^{-1/2-i\mu}
&=
\frac{1}{k_3^3}
\left(\frac{k_3}{K}\right)^{5/2}
\left(\frac{k_3}{K}\right)^{-i\mu},
\\
K^{-5/2-i\mu}k_3^{-1/2+i\mu}
&=
\frac{1}{k_3^3}
\left(\frac{k_3}{K}\right)^{5/2}
\left(\frac{k_3}{K}\right)^{+i\mu}.
\end{align}
These are the two conjugate clock signals. The final Schwinger-Keldysh combination relevant for the exchange diagram is obtained after including the relative signs of the Schwinger-Keldysh vertices:
\begin{align}
\mathcal I_{\rm SK}
\equiv
\mathcal I^{++}_{12|3}
+
\mathcal I^{--}_{12|3}
-
\mathcal I^{+-}_{12|3}
-
\mathcal I^{-+}_{12|3}.
\label{app:eq:ISK_def}
\end{align}
The soft contribution therefore takes the structural form
\begin{align}
\mathcal C_{12|3}^{\rm non-an.}
=
\frac{H}{4k_1k_2k_3}
\left[
\mathcal S_-(\mu,\lambda,\gamma)
K^{-5/2+i\mu}k_3^{-1/2-i\mu}
+
\mathcal S_+(\mu,\lambda,\gamma)
K^{-5/2-i\mu}k_3^{-1/2+i\mu}
\right],
\end{align}
up to terms analytic in \(k_3/K\). For real parameters, \(\mathcal S_+=\mathcal S_-^*\). The explicit form of \(\mathcal S_\pm\) is obtained by applying the monomial rules above to each term in \eqref{app:eq:GLT_decomp}.

\section{Hard channels}
\label{app:hard_channels}

The two remaining exchange channels are $\mathcal C_{23|1}$ and $\mathcal C_{31|2}$. In the squeezed limit, these are hard channels because $s_{23}=k_1$ and $s_{31}=k_2$, which remain finite as \(k_3\to0\). Therefore, one should not use the small-\(s\) expansion of the internal propagator in these channels. For the first hard channel,
\begin{align}
\mathcal C_{23|1}
=
\frac{H}{4k_1k_2k_3}
\sum_{\epsilon,\epsilon'=\pm}
\mathcal I^{\epsilon\epsilon'}_{23|1},
\end{align}
where
\begin{align}
\mathcal I^{\epsilon\epsilon'}_{23|1}
=
\int_{-\infty}^{0}d\eta_1
\int_{-\infty}^{0}
\frac{d\eta_2}{(-\eta_2)^2}\,
e^{i\epsilon(k_2+k_3)\eta_1+i\epsilon'k_1\eta_2}
G_{\epsilon\epsilon'}(k_1,\eta_1,\eta_2).
\label{app:eq:I_hard_231}
\end{align}
Similarly,
\begin{align}
\mathcal I^{\epsilon\epsilon'}_{31|2}
=
\int_{-\infty}^{0}d\eta_1
\int_{-\infty}^{0}
\frac{d\eta_2}{(-\eta_2)^2}\,
e^{i\epsilon(k_3+k_1)\eta_1+i\epsilon'k_2\eta_2}
G_{\epsilon\epsilon'}(k_2,\eta_1,\eta_2).
\label{app:eq:I_hard_312}
\end{align}
Let us focus on \(\mathcal C_{23|1}\). We define $x\equiv -k_1\eta_1$ and $y\equiv -k_1\eta_2$. Then $ d\eta_1\,{d\eta_2}/{(-\eta_2)^2} ={dx\,dy}/{y^2}$, and $ {i\epsilon(k_2+k_3)\eta_1+i\epsilon'k_1\eta_2}={-i\epsilon r x-i\epsilon' y}$, and $r\equiv {(k_2+k_3)}/{k_1}$. Thus,
\begin{align}
\mathcal C_{23|1}
=
\frac{H}{4k_1k_2k_3}
\sum_{\epsilon,\epsilon'=\pm}
\int_0^\infty dx
\int_0^\infty \frac{dy}{y^2}\,
e^{-i\epsilon r x-i\epsilon' y}
G_{\epsilon\epsilon'}(x,y;\lambda,\gamma).
\label{app:eq:hard_rescaled}
\end{align}

In the squeezed limit, $ r=1+\mathcal O\!\left({k_3}/{k_1}\right)$. Expanding the external phase,
\begin{align}
e^{-i\epsilon r x}
=
e^{-i\epsilon x}
\sum_{n=0}^{\infty}
\frac{(-i\epsilon x)^n}{n!}
(r-1)^n,
\end{align}
one obtains an analytic expansion of the hard channel:
\begin{align}
\mathcal C_{23|1}
\simeq
\sum_{n=0}^{\infty}
C_n^{(23)}(\mu,\lambda,\gamma)
\left(\frac{k_3}{k_1}\right)^n,
\end{align}
where, schematically,
\begin{align}
C_n^{(23)}(\mu,\lambda,\gamma)
=
\frac{H}{4k_1k_2k_3}
\frac{1}{n!}
\sum_{\epsilon,\epsilon'=\pm}
(-i\epsilon)^n
\int_0^\infty dx
\int_0^\infty\frac{dy}{y^2}\,
x^n
e^{-i\epsilon x-i\epsilon' y}
G_{\epsilon\epsilon'}(x,y;\lambda,\gamma).
\label{app:eq:Cn_general}
\end{align}
The second hard channel gives the same structure, with the exchange \(k_1\leftrightarrow k_2\). In the symmetric squeezed configuration \(k_1\simeq k_2\), the two hard channels contribute equally; therefore,
\begin{align}
\big\langle
\varphi_{\mathbf k_1}
\varphi_{\mathbf k_2}
\varphi_{\mathbf k_3}
\big\rangle'_{\rm hard}
\equiv
\mathcal C_{23|1}+\mathcal C_{31|2}
\simeq
2
\sum_{n=0}^{\infty}
C_n(\mu,\lambda,\gamma)
\left(\frac{k_3}{k_1}\right)^n .
\label{app:eq:hard_series}
\end{align}
This contribution is analytic in the squeezed ratio \(k_3/k_1\). It is
therefore a background contribution rather than a non-analytic clock signal. For completeness, one may write the hard coefficients directly in terms of the exact Whittaker mode functions. Define
\begin{align}
\mathscr W_+(x)
&\equiv
W_{-i\lambda,\,i\mu/2}\!\left(i\gamma x^2\right)
\qquad , \qquad 
\mathscr W_-(x)
\equiv
W_{+i\lambda,\,-i\mu/2}\!\left(-i\gamma x^2\right)
\quad .
\end{align}
Then the hard coefficients can be organized as:
\begin{align}
C_n(\mu,\lambda,\gamma)
&=
\frac{H^3}{8\gamma k_1^5k_3}
\frac{1}{n!}
\int_0^\infty dx
\int_0^\infty\frac{dy}{y^2}\,
x^n(xy)^{1/2}
\Bigg\{
\nonumber\\
&\quad
(-i)^n e^{-ix-iy}
\Big[
\Theta(y-x)\mathscr W_+(x)\mathscr W_-(y)
+
\Theta(x-y)\mathscr W_-(x)\mathscr W_+(y)
\Big]
\nonumber\\
&\quad
+
(-i)^n e^{-ix+iy}
\mathscr W_-(x)\mathscr W_+(y)
+
(i)^n e^{ix-iy}
\mathscr W_+(x)\mathscr W_-(y)
\nonumber\\
&\quad
+
(i)^n e^{ix+iy}
\Big[
\Theta(x-y)\mathscr W_+(x)\mathscr W_-(y)
+
\Theta(y-x)\mathscr W_-(x)\mathscr W_+(y)
\Big]
\Bigg\}.
\label{app:eq:Cn_whittaker}
\end{align}
If the Wightman propagator carries a universal normalization factor \(e^{-\pi\lambda}\), independent of the Schwinger-Keldysh branch indices and the integration variables, then one may factor it out as
\begin{align}
C_n(\mu,\lambda,\gamma)
=
e^{-\pi\lambda}
\widehat C_n(\mu,\lambda,\gamma).
\end{align}
This factorization does not remove the dependence on the tilted dispersion relation. The remaining dependence on \(\lambda\) and \(\gamma\) is still present inside the Whittaker indices and arguments. The factor \(e^{-\pi\lambda}\) therefore controls the overall weight of the hard contribution, while the detailed shape of the analytic coefficients is fixed by the exact internal propagator.

\section{Whittaker modes and useful limits}
\label{subsec:soft_whittaker_corrected}

In this appendix we show explicitly how the tilted-ghost mode equation can be reduced to the standard Whittaker equation. We start from the canonical spectator mode equation
\begin{equation}
f_k''(\eta)
+
\left[
s_{\rm tilt} \beta k^2
+
\gamma^2 k^4\eta^2
-
\frac{\nu^2-\frac14}{\eta^2}
\right]f_k(\eta)=0,
\qquad
\nu=\sqrt{\frac94-\frac{m^2}{H^2}},
\label{eq:app_tilted_mode}
\end{equation}
where $ s_{\rm tilt}=+1$ corresponds to the negative-tilt branch and $s_{\rm tilt}=-1$ corresponds to the positive-tilt branch. We also introduce the signed tilt parameter
\begin{equation}
\lambda \equiv s_{\rm tilt}\frac{\beta}{4\gamma}.
\label{eq:app_signed_lambda}
\end{equation}
Thus the negative-tilt branch has $\lambda=+\beta/(4\gamma)$, whereas the positive-tilt branch has $\lambda=-\beta/(4\gamma)$.

We now define
\begin{equation}
z=i\gamma k^2\eta^2,
\qquad
f_k(\eta)=g(z),
\label{eq:app_z_definition}
\end{equation}
so that primes on $g$ denote derivatives with respect to $z$. Since
\begin{equation}
\frac{d f_k}{d\eta}
=
2i\gamma k^2\eta,g'(z),
\qquad
\frac{d^2 f_k}{d\eta^2}
=
2i\gamma k^2 g'(z)
+
4i\gamma k^2 z,g''(z),
\end{equation}
Eq.~\eqref{eq:app_tilted_mode} becomes
\begin{equation}
z g''(z)
+
\frac12 g'(z)
+
\left[
-\frac{z}{4}
-
i\lambda
-
\frac{\nu^2-\frac14}{4z}
\right]g(z)=0.
\label{eq:app_g_equation}
\end{equation}
The first-derivative term can be removed by writing
\begin{equation}
g(z)=z^{-1/4}h(z).
\label{eq:app_g_to_h}
\end{equation}
Indeed, this substitution gives
\begin{equation}
z g''(z)+\frac12 g'(z)
=
z^{3/4}h''(z)
+
\frac{3}{16}z^{-5/4}h(z),
\end{equation}
and therefore Eq.~\eqref{eq:app_g_equation} reduces to
\begin{equation}
h''(z)
+
\left[
-\frac14
-
\frac{i\lambda}{z}
+
\frac{1-\nu^2}{4z^2}
\right]h(z)=0.
\label{eq:app_whittaker_form}
\end{equation}
This is the standard Whittaker differential equation,
\begin{equation}
h''(z)
+
\left[
-\frac14
+
\frac{\kappa}{z}
+
\frac{\frac14-\alpha^2}{z^2}
\right]h(z)=0,
\label{eq:app_standard_whittaker}
\end{equation}
with the identification
\begin{equation}
\kappa=-i\lambda,
\qquad
\alpha=\frac{\nu}{2}.
\label{eq:app_whittaker_parameters}
\end{equation}
The two independent solutions may therefore be written as
\begin{equation}
h(z)=c_M,M_{-i\lambda,\nu/2}(z)
+
c_W,W_{-i\lambda,\nu/2}(z),
\end{equation}
and hence
\begin{equation}
f_k(\eta)
=
z^{-1/4}
\left[
c_M,M_{-i\lambda,\nu/2}(z)
+
c_W,W_{-i\lambda,\nu/2}(z)
\right],
\qquad
z=i\gamma k^2\eta^2.
\label{eq:app_general_whittaker_solution}
\end{equation}
The branch relevant for the tilted-ghost vacuum is selected by the early-time behavior. The large-$|z|$ asymptotic form of the Whittaker-$W$ function is
\begin{equation}
W_{\kappa,\alpha}(z)
\sim
e^{-z/2}z^\kappa,
\qquad
|z|\rightarrow\infty,
\label{eq:app_large_z_whittaker}
\end{equation}
inside the appropriate Stokes wedge. Using $z=i\gamma k^2\eta^2$, this reproduces the pure-ghost WKB phase
\begin{equation}
\exp\left(-\frac{i}{2}\gamma k^2\eta^2\right),
\end{equation}
up to the slowly varying phase associated with the subleading quadratic term. The normalized positive-frequency branch used in the main text is therefore
\begin{equation}
f_k(\eta)
=
-
\frac{e^{-\pi\lambda/2}}
{\sqrt{2}\sqrt{\gamma},k,(-\eta)^{1/2}}
W_{-i\lambda,\nu/2}
\left(i\gamma k^2\eta^2\right).
\label{eq:app_normalized_whittaker_mode}
\end{equation}
The physical spectator fluctuation is obtained from
\begin{equation}
\sigma_k(\eta)=\frac{f_k(\eta)}{a(\eta)}
=
-H\eta,f_k(\eta).
\end{equation}
Finally, the late-time behavior follows from the Whittaker connection formula
\begin{equation}
W_{\kappa,\alpha}(z)
=
\frac{\Gamma(-2\alpha)}
{\Gamma\left(\frac12-\alpha-\kappa\right)}
M_{\kappa,\alpha}(z)
+
\frac{\Gamma(2\alpha)}
{\Gamma\left(\frac12+\alpha-\kappa\right)}
M_{\kappa,-\alpha}(z),
\label{eq:app_connection_formula}
\end{equation}
together with
\begin{equation}
M_{\kappa,\alpha}(z)
\sim
z^{\alpha+1/2},
\qquad
z\rightarrow0.
\end{equation}
Therefore,
\begin{equation}
W_{\kappa,\alpha}(z)
\sim
\frac{\Gamma(-2\alpha)}
{\Gamma\left(\frac12-\alpha-\kappa\right)}
z^{1/2+\alpha}
+
\frac{\Gamma(2\alpha)}
{\Gamma\left(\frac12+\alpha-\kappa\right)}
z^{1/2-\alpha},
\qquad
z\rightarrow0.
\label{eq:app_small_z_whittaker}
\end{equation}
For principal-series fields, $\nu=i\mu$, this gives the two late-time oscillatory branches
\begin{equation}
\sigma_k(\eta)
\sim
A_+(k,\mu,\lambda,\gamma)(-\eta)^{3/2+i\mu}
+
A_-(k,\mu,\lambda,\gamma)(-\eta)^{3/2-i\mu},
\end{equation}
with the branch weights fixed by the Whittaker connection coefficients. This is the origin of the $\lambda$-dependent asymmetry between the two cosmological-collider clock branches.

The Schwinger-Keldysh calculation in the soft channel is already contained in the monomial
rules above.  The only extra input needed for the explicit Whittaker comparison
is the late-time branch decomposition of the internal propagator.  For the
standard Whittaker normalization used in the boostless template,
\begin{align}
\sigma_k(\eta_1)\sigma_k^*(\eta_2)
&=
\frac{H^2\,e^{-\pi\lambda}}{2 c_b k }\,
\eta_1\eta_2 
\mathcal{W}_{i\lambda,\, i\mu/2}\!\big(2ic_b k \eta_1\big) \mathcal{W}_{-i\lambda,\,-i\mu/2}\!\big(-2 i c_b k \eta_2 \big),
\label{standard_whitakker_modes}
\end{align}
we write
\begin{align}
\mathcal G^{\rm W}_{\rm LT}(k;\eta_1,\eta_2)
&=
\widehat{\mathcal A}_{-}
 k^{-2i\mu}
(-\eta_1)^{\Delta_-}(-\eta_2)^{\Delta_-}
+
\widehat{\mathcal B}_{-+}
(-\eta_1)^{\Delta_-}(-\eta_2)^{\Delta_+}
\nonumber\\
&\quad+
\widehat{\mathcal A}_{+}
 k^{2i\mu}
(-\eta_1)^{\Delta_+}(-\eta_2)^{\Delta_+}
+
\widehat{\mathcal B}_{+-}
(-\eta_1)^{\Delta_+}(-\eta_2)^{\Delta_-},
\label{app:eq:standard_whittaker_LT_summary}
\end{align}
The coefficients are
\begin{align}
\widehat{\mathcal A}_{-}
&=
H^2e^{-\pi\lambda}(2c_b)^{-2i\mu}
\frac{\Gamma(2i\mu)^2}
{\Gamma\!\left(\frac12-i\lambda+i\mu\right)
 \Gamma\!\left(\frac12+i\lambda+i\mu\right)},
\nonumber\\
\widehat{\mathcal A}_{+}
&=
H^2e^{-\pi\lambda}(2c_b)^{2i\mu}
\frac{\Gamma(-2i\mu)^2}
{\Gamma\!\left(\frac12-i\lambda-i\mu\right)
 \Gamma\!\left(\frac12+i\lambda-i\mu\right)},
\nonumber\\
\widehat{\mathcal B}_{-+}
&=
\frac{H^2}{2\mu}
 e^{-\pi\lambda-\pi\mu}
\csch(2\pi\mu)
\cosh\!\left[\pi(\lambda-\mu)\right],
\nonumber\\
\widehat{\mathcal B}_{+-}
&=
\frac{H^2}{2\mu}
 e^{-\pi\lambda+\pi\mu}
\csch(2\pi\mu)
\cosh\!\left[\pi(\lambda+\mu)\right].
\label{app:eq:standard_whittaker_coefficients_summary}
\end{align}
This is the only point at which the standard Whittaker template differs from
the tilted-ghost calculation of the main text: the Schwinger-Keldysh phases, the ordered
integrals, and the soft scaling rules are identical.  One simply replaces the
late-time coefficients \((\mathcal A_\pm,\mathcal B_{-+},\mathcal B_{+-})\) in
Eq.~\eqref{app:eq:GLT_decomp} with the hatted coefficients in
Eq.~\eqref{app:eq:standard_whittaker_coefficients_summary}. Therefore, the
non-analytical part is obtained immediately from the same Schwinger-Keldysh combination.
\begin{align}
\mathcal I_{{\rm W},{\rm non\mbox{-}an}}
&\simeq
\frac{2H^2e^{-\pi\lambda}}{k_3^3}
\left(\frac{k_3}{K}\right)^{5/2}
\left(4c_b^2\frac{k_3}{K}\right)^{-i\mu}
\nonumber\\
&\quad\times
\frac{
\bigl[1-i\sinh(\pi\mu)\bigr]
\Gamma\!\left(\frac12-i\mu\right)
\Gamma\!\left(\frac52-i\mu\right)
\Gamma(2i\mu)^2
}{
\Gamma\!\left(\frac12-i\lambda+i\mu\right)
\Gamma\!\left(\frac12+i\lambda+i\mu\right)
}
+
{\rm c.c.}
\label{app:eq:standard_whittaker_nonlocal_summary}
\end{align}
The Hankel limit follows by taking \(\lambda\to0\) and using
\begin{align}
\Gamma(2i\mu)
=
2^{2i\mu-1}\pi^{-1/2}
\Gamma(i\mu)
\Gamma\!\left(\frac12+i\mu\right).
\end{align}
This gives
\begin{align}
\mathcal I_{{\rm H},{\rm non\mbox{-}an}}
&\simeq
\frac{H^2}{2\pi k_3^3}
\left(\frac{k_3}{K}\right)^{5/2}
\left(\frac{c_b^2 k_3}{4K}\right)^{-i\mu}
\nonumber\\
&\quad\times
\bigl[1-i\sinh(\pi\mu)\bigr]
\Gamma\!\left(\frac12-i\mu\right)
\Gamma\!\left(\frac52-i\mu\right)
\Gamma(i\mu)^2
+
{\rm c.c.}
\label{app:eq:standard_hankel_nonlocal_summary}
\end{align}
Thus the standard Whittaker result reduces to the Hankel expression when the
Whittaker deformation is removed.  By contrast, the tilted-ghost Whittaker
result reduces to the pure-ghost limit shown in
Eq.~\eqref{eq:TG_pure_ghost_limit_main}.  The two limits have the same clock
exponent but differ in their normalization and in the parameter that controls
the UV branch: \(c_b\) for the standard template and \(\gamma\) for the
tilted-ghost/pure-ghost mode.

\subsection{Large-mass limit}
\label{app:large-mu-whittaker}

We now record the large-$\mu$ behavior of the Whittaker connection coefficients used in the main text. This limit is useful for understanding the large-mass behavior of the tilted-ghost branch asymmetry. The relevant small-$z$ connection formula is
\begin{equation}
W_{\kappa,\alpha}(z)
\simeq
\frac{\Gamma(-2\alpha)}
{\Gamma\left(\frac12-\alpha-\kappa\right)}
z^{1/2+\alpha}
+
\frac{\Gamma(2\alpha)}
{\Gamma\left(\frac12+\alpha-\kappa\right)}
z^{1/2-\alpha},
\qquad
z\rightarrow 0 .
\end{equation}
For the tilted-ghost mode, $\kappa=-i\lambda$ and $\alpha=i\mu/2$. Therefore,
\begin{equation}
W_{-i\lambda,i\mu/2}(z)
\simeq
{\cal C}*{+}(\mu,\lambda),z^{1/2+i\mu/2}
+
{\cal C}*{-}(\mu,\lambda),z^{1/2-i\mu/2},
\end{equation}
where
\begin{equation}
{\cal C}*{+}(\mu,\lambda)
=
\frac{\Gamma(-i\mu)}
{\Gamma\left(\frac12+i\lambda-\frac{i\mu}{2}\right)},
\qquad
{\cal C}*{-}(\mu,\lambda)
=
\frac{\Gamma(i\mu)}
{\Gamma\left(\frac12+i\lambda+\frac{i\mu}{2}\right)} .
\label{eq:app_connection_coefficients_large_mu}
\end{equation}
Using Stirling's approximation for fixed $\lambda$ and $\mu\gg |\lambda|$, the absolute values become
\begin{equation}
|{\cal C}*{+}(\mu,\lambda)|
\simeq
\mu^{-1/2}
\exp\left[
-\frac{\pi\mu}{4}
-
\frac{\pi\lambda}{2}
\right]
\left[
1+{\cal O}\left(\frac{1}{\mu}\right)
\right],
\label{eq:app_Cplus_large_mu}
\end{equation}
and
\begin{equation}
|{\cal C}*{-}(\mu,\lambda)|
\simeq
\mu^{-1/2}
\exp\left[
-\frac{\pi\mu}{4}
+
\frac{\pi\lambda}{2}
\right]
\left[
1+{\cal O}\left(\frac{1}{\mu}\right)
\right].
\label{eq:app_Cminus_large_mu}
\end{equation}
Hence the relative branch weight approaches
\begin{equation}
\frac{|{\cal C}*{+}|}{|{\cal C}*{-}|}
\simeq
e^{-\pi\lambda},
\qquad
\frac{|{\cal C}*{+}|^2}{|{\cal C}*{-}|^2}
\simeq
e^{-2\pi\lambda},
\qquad
\mu\gg |\lambda|.
\label{eq:app_large_mu_branch_ratio}
\end{equation}
This reproduces the large-$\mu$ limit of the exact branch ratio,
\begin{equation}
\frac{|{\cal C}*{+}|^2}{|{\cal C}*{-}|^2}
=
\frac{\cosh\left[\pi\left(\lambda-\frac{\mu}{2}\right)\right]}
{\cosh\left[\pi\left(\lambda+\frac{\mu}{2}\right)\right]}
\longrightarrow
e^{-2\pi\lambda}.
\end{equation}
Thus, after the common large-$\mu$ suppression is factored out, the remaining branch asymmetry is controlled by the signed tilt $\lambda$.

The leading phases can also be obtained from Stirling's approximation. Writing
\begin{equation}
{\cal C}*{\pm}(\mu,\lambda)
=
|{\cal C}*{\pm}(\mu,\lambda)|
e^{i\Theta_{\pm}(\mu,\lambda)},
\end{equation}
one finds, for $\mu\gg|\lambda|$,
\begin{equation}
\Theta_{+}(\mu,\lambda)
\simeq
\left(\frac{\mu}{2}-\lambda\right)
\left[
\log\left(\frac{\mu}{2}-\lambda\right)-1
\right]
-
\mu(\log\mu-1)
+
\frac{\pi}{4},
\label{eq:app_theta_plus_large_mu}
\end{equation}
and
\begin{equation}
\Theta_{-}(\mu,\lambda)
\simeq
\mu(\log\mu-1)
-
\left(\frac{\mu}{2}+\lambda\right)
\left[
\log\left(\frac{\mu}{2}+\lambda\right)-1
\right]
-
\frac{\pi}{4}.
\label{eq:app_theta_minus_large_mu}
\end{equation}
The moduli in Eqs.~\eqref{eq:app_Cplus_large_mu} and \eqref{eq:app_Cminus_large_mu} explain the large-mass plateau of the normalized amplitude ratios, while the phases in Eqs.~\eqref{eq:app_theta_plus_large_mu} and \eqref{eq:app_theta_minus_large_mu} determine the corresponding displacement of the logarithmic clock signal.

\subsection{Cosmological collider form}

We start from the squeezed-limit Schwinger-Keldysh integral
\begin{align*}
\mathcal I_{\rm SK}
&\simeq
2\Gamma\!\left(\frac52-i\mu\right)
K^{-5/2+i\mu}
k_3^{-1/2-i\mu}
\Bigg[
\left(1-i\sinh\pi\mu\right)
\widehat{\mathcal A}_-
\Gamma\!\left(\frac12-i\mu\right)
+
\widehat{\mathcal B}^{{\rm SK}}_-
\Gamma\!\left(\frac12+i\mu\right)
\Bigg]
\nonumber\\
&+
2\Gamma\!\left(\frac52+i\mu\right)
K^{-5/2-i\mu}
k_3^{-1/2+i\mu}
\Bigg[
\left(1+i\sinh\pi\mu\right)
\widehat{\mathcal A}_+
\Gamma\!\left(\frac12+i\mu\right)
+
\widehat{\mathcal B}^{{\rm SK}}_+
\Gamma\!\left(\frac12-i\mu\right)
\Bigg].
\end{align*}
We set \(K=2k_1\) and define the corresponding shape contribution by $S_{\rm SK}
\equiv \mathcal{N} k_1^2k_3\,\mathcal I_{\rm SK}$. After multiplying by \(k_1^2k_3\), the two momentum structures become
\begin{align}
k_1^2k_3\,
(2k_1)^{-5/2+i\mu}
k_3^{-1/2-i\mu}
&=
2^{-5/2+i\mu}
k_1^{-1/2+i\mu}
k_3^{1/2-i\mu},
\\
k_1^2k_3\,
(2k_1)^{-5/2-i\mu}
k_3^{-1/2+i\mu}
&=
2^{-5/2-i\mu}
k_1^{-1/2-i\mu}
k_3^{1/2+i\mu}.
\end{align}
Then the shape can be written as
\begin{equation}
\boxed{
S_{\rm SK}
\simeq
\frac{1}{2}
\left(\frac{k_3}{k_1}\right)^{1/2}
\left[
\mathcal C_-
\left(\frac{c_b^2 k_3}{k_1}\right)^{-i\mu}
+
\mathcal C_+
\left(\frac{c_b^2 k_3}{k_1}\right)^{i\mu}
\right].
}
\label{eq:SK_shape_collider_complex}
\end{equation}
With this convention, all numerical factors have been absorbed into the complex coefficients. The positive-frequency coefficient is
\begin{align}
\boxed{
\begin{aligned}
\mathcal C_+
&=
2^{-1/2+i\mu}\,
e^{-\pi \lambda}
\left(1+i\sinh\pi\mu\right)
\frac{
\Gamma\!\left(\frac52+i\mu\right)
\Gamma\!\left(\frac12+i\mu\right)
\Gamma(-2i\mu)^2
}{
\Gamma\!\left(\frac12-i \lambda-i\mu\right)
\Gamma\!\left(\frac12+i \lambda-i\mu\right)
}
\nonumber\\[4pt]
&\quad
+
\frac{2^{-5/2-i\mu}}{\mu}\,
c_b^{-2i\mu}\,
\Gamma\!\left(\frac52+i\mu\right)
\Gamma\!\left(\frac12-i\mu\right)
\left[
\frac12
\left(1+e^{-2\pi \lambda}\right)
\csch(\pi\mu)
+i
\right].
\end{aligned}
}
\label{eq:Cplus_kappa_final}
\end{align}
Similarly,
\begin{align}
\boxed{
\begin{aligned}
\mathcal C_-
&=
2^{-1/2-i\mu}\,
e^{-\pi \lambda}
\left(1-i\sinh\pi\mu\right)
\frac{
\Gamma\!\left(\frac52-i\mu\right)
\Gamma\!\left(\frac12-i\mu\right)
\Gamma(2i\mu)^2
}{
\Gamma\!\left(\frac12-i \lambda+i\mu\right)
\Gamma\!\left(\frac12+i \lambda+i\mu\right)
}
\nonumber\\[4pt]
&\quad
+
\frac{2^{-5/2+i\mu}}{\mu}\,
c_b^{2i\mu}\,
\Gamma\!\left(\frac52-i\mu\right)
\Gamma\!\left(\frac12+i\mu\right)
\left[
\frac12
\left(1+e^{-2\pi \lambda}\right)
\csch(\pi\mu)
-i
\right].
\end{aligned}
}
\end{align}
For real \(\mu\), \(\lambda\), and \(c_b\), the two coefficients obey
$\mathcal C_- =\left(\mathcal C_+ \right)^\ast$. Therefore, the Schwinger-Keldysh contribution can be written in the real cosmological-collider form
\begin{equation}
\boxed{
S_{\rm SK}
\simeq
\left(\frac{k_3}{k_1}\right)^{1/2}
\left|\mathcal C_+ \right|
\cos\left[
\mu\log\left(
\frac{c_b^2 k_3}{k_1}
\right)
+
\arg\mathcal C_+
\right].
}
\label{eq:SK_shape_collider_real}
\end{equation}
Thus, the squeezed-limit signal has the standard collider structure: an envelope $
\left({k_3}/{k_1}\right)^{1/2}$, multiplying a logarithmic oscillation with phase $\mu\log\left( \frac{c_b^2 k_3}{k_1} \right)$. The tilt parameter \(\lambda\) enters through both the amplitude
\(\left|\mathcal C_+\right|\) and the residual phase \(\arg\mathcal C_+\).

\bibliographystyle{JHEP}
\bibliography{biblio}

\end{document}